\documentclass[twocolumn,longauth]{aa}

\usepackage[table]{xcolor}
\usepackage{graphicx}
\usepackage[version=4]{mhchem}
\usepackage{txfonts}
\usepackage{orcidlink}
\usepackage{academicons}
\usepackage{tablefootnote}
\usepackage{gensymb}
\usepackage[normalem]{ulem}
\usepackage{ifsym}
\usepackage{caption,subcaption}

\definecolor{red}{rgb}{0.7, 0.11, 0.11}
\definecolor{yellow}{rgb}{1.0, 0.75, 0.0}
\definecolor{green}{rgb}{0.53, 0.66, 0.42}

\makeatletter
\renewcommand*\aa@pageof{, page \thepage{} of \pageref*{LastPage}}

\begin{document}

\title{MINDS. The influence of outer dust disc structure on the volatile delivery to the inner disc}

\author{Danny Gasman
        \inst{1}*\orcidlink{0000-0002-1257-7742}
        \and
        Milou Temmink
        \inst{2}\orcidlink{0000-0002-7935-7445}
        \and
        Ewine F. van Dishoeck
        \inst{2,3}\orcidlink{0000-0001-7591-1907}
        \and
        Nicolas T. Kurtovic
        \inst{3}\orcidlink{0000-0002-2358-4796}
        \and
        Sierra L. Grant
        \inst{3}\orcidlink{0000-0002-4022-4899}
        \and
        Andrew Sellek
        \inst{2}\orcidlink{0000-0003-0330-1506}
        \and
        Beno\^{i}t Tabone
        \inst{4}
        \and
        Thomas Henning
        \inst{5}\orcidlink{0000-0002-1493-300X}
        \and
        Inga Kamp
        \inst{6}\orcidlink{0000-0001-7455-5349}
        Manuel G\"udel
        \inst{7,8}\orcidlink{0000-0001-9818-0588}
        \and
        David Barrado
        \inst{9}\orcidlink{0000-0002-5971-9242}
        \and
        Alessio Caratti o Garatti
        \inst{10,11}\orcidlink{0000-0001-8876-6614}
        \and
        Adrian M. Glauser
        \inst{8}\orcidlink{0000-0001-9250-1547}
        \and
        L. B. F. M. Waters
        \inst{12,13}\orcidlink{0000-0002-5462-9387}
        \and
        Aditya M. Arabhavi
        \inst{6}\orcidlink{0000-0001-8407-4020}
        \and
        Hyerin Jang
        \inst{12}\orcidlink{0000-0002-6592-690X}
        \and
        Jayatee Kanwar
        \inst{6,14,15}\orcidlink{0000-0003-0386-2178}
        \and
        Julia L. Lienert
        \inst{5}
        \and
        Giulia Perotti
        \inst{5,16}\orcidlink{0000-0002-8545-6175}
        \and
        Kamber Schwarz
        \inst{5}\orcidlink{0000-0002-6429-9457}
        \and
        Marissa Vlasblom
        \inst{2}\orcidlink{0000-0002-3135-2477}}
        
\institute{Institute of Astronomy, Celestijnenlaan 200D, 3001, Leuven, Belgium
        \and
        Leiden Observatory, Leiden University, 2300 RA Leiden, the Netherlands
        \and
        Max-Planck Institut f\"{u}r Extraterrestrische Physik (MPE), Giessenbachstr. 1, 85748, Garching, Germany
        \and
        Universit\'e Paris-Saclay, CNRS, Institut d’Astrophysique Spatiale, 91405, Orsay, France
        \and
        Max-Planck-Institut f\"{u}r Astronomie (MPIA), K\"{o}nigstuhl 17, 69117 Heidelberg, Germany
        \and
        Kapteyn Astronomical Institute, Rijksuniversiteit Groningen, Postbus 800, 9700AV Groningen, The Netherlands
        \and
        Dept. of Astrophysics, University of Vienna, T\"urkenschanzstr. 17, A-1180 Vienna, Austria
        \and
        ETH Z\"urich, Institute for Particle Physics and Astrophysics, Wolfgang-Pauli-Str. 27, 8093 Z\"urich, Switzerland
        \and
        Centro de Astrobiolog\'ia (CAB), CSIC-INTA, ESAC Campus, Camino Bajo del Castillo s/n, 28692 Villanueva de la Ca\~nada, Madrid, Spain
        \and
        INAF – Osservatorio Astronomico di Capodimonte, Salita Moiariello 16, 80131 Napoli, Italy
        \and
		Dublin Institute for Advanced Studies, 31 Fitzwilliam Place, D02 XF86 Dublin, Ireland
		\and
		Department of Astrophysics/IMAPP, Radboud University, PO Box 9010, 6500 GL Nijmegen, The Netherlands
		\and
		SRON Netherlands Institute for Space Research, Niels Bohrweg 4, NL-2333 CA Leiden, the Netherlands
		\and
		Space Research Institute, Austrian Academy of Sciences, Schmiedlstr. 6, A-8042, Graz, Austria
		\and
		TU Graz, Fakultät für Mathematik, Physik und Geodäsie, Petersgasse 16 8010 Graz, Austria
		\and
		Niels Bohr Institute, University of Copenhagen, NBB BA2, Jagtvej 155A, 2200 Copenhagen, Denmark
        \\
            *\email{danny.gasman@kuleuven.be}}

   \date{Received 9 September 2024 / Accepted 20 December 2024}

  \abstract
   {The Atacama Large Millimeter/submillimeter Array (ALMA) has revealed that the millimetre dust structures of protoplanetary discs are extremely diverse, ranging from small and compact dust discs to large discs with multiple rings and gaps. It has been proposed that the strength of \ce{H2O} emission in the inner disc particularly depends on the influx of icy pebbles from the outer disc, a process that would correlate with the outer dust disc radius, and that could be prevented by pressure bumps. Additionally, the dust disc structure should also influence the emission of other gas species in the inner disc. Since terrestrial planets likely form in the inner disc regions, understanding their composition is of interest.}
   {This work aims to assess the influence of pressure bumps on the inner disc's molecular reservoirs. The presence of a dust gap, and potentially giant planet formation farther out in the disc, may influence the composition of the inner disc, and thus the building blocks of terrestrial planets.}
   {Using the improved sensitivity and spectral resolution of the Mid-InfraRed Instrument's (MIRI) Medium Resolution Spectrometer (MRS) on the \textit{James Webb} Space Telescope (\textit{JWST}) compared to \textit{Spitzer}, we compared the observational emission properties of \ce{H2O}, \ce{HCN}, \ce{C2H2}, and \ce{CO2} with the outer dust disc structure from ALMA observations, in eight discs with confirmed gaps in ALMA observations, and two discs with gaps of tens of astronomical units in width, around stars with $M_\star \geq 0.45M_{\odot}$. We used new visibility plane fits of the ALMA data to determine the outer dust disc radius and identify substructures in the discs.}
   {We find that the presence of a dust gap does not necessarily result in weak \ce{H2O} emission. Furthermore, the relative lack of colder \ce{H2O}-emission seems to go hand in hand with elevated emission from carbon-bearing species. Of the discs that show significant substructure within the \ce{CO} and \ce{CH4} snowlines, most show detectable emission from the carbon-bearing species. The discs with cavities and extremely wide gaps appear to behave as a somewhat separate group, with stronger cold \ce{H2O} emission and weak warm \ce{H2O} emission.}
   {We conclude that fully blocking radial dust drift from the outer disc seems difficult to achieve, even for discs with very wide gaps or cavities, which can still show significant cold \ce{H2O} emission. However, there does seem to be a dichotomy between discs that show a strong cold \ce{H2O} excess and ones that show strong emission from \ce{HCN} and \ce{C2H2}. Better constraints on the influence of the outer dust disc structure and inner disc composition require more information on substructure formation timescales and disc ages, along with the importance of trapping of (hyper)volatiles like CO and \ce{CO2} into more strongly bound ices like \ce{H2O} and chemical transformation of CO into less volatile species.}


    \titlerunning{The influence of outer dust disc structure on the volatile delivery to the inner disc}
    \authorrunning{D. Gasman et al.}
   \maketitle

\section{Introduction} \label{sec:intro}

The inner 10~au of protoplanetary discs observable with the Medium Resolution Spectrometer \citep[MRS;][]{ref:15WePeGl,ref:23ArGlLa} of the Mid-InfraRed Instrument \citep[MIRI;][]{ref:15WrWrGo,ref:15RiWrBo,ref:23WrRiGl} on board the \textit{James Webb} Space Telescope \citep[\textit{JWST};][]{ref:23RiPeMc} is known to host active chemistry due to high temperatures and densities. This part of discs is of particular interest, as it is likely where terrestrial planets form \citep[e.g.][]{ref:21ObBe,ref:22MoMoBi}. This region has been studied previously with \textit{Spitzer}'s InfraRed Spectrograph (IRS) \citep[e.g.][]{ref:09PaApLu,ref:10PoSaBl,ref:11CaNa,ref:13PaHeCa,ref:14PoSaBe}, and ground-based data \citep[e.g.][]{ref:13BrPoDi,ref:22BaAbBr,ref:23BaPoPe}, but the launch of \textit{JWST} allows us to re-evaluate these regions with an improved sensitivity and spectral resolution \citep[e.g.][]{ref:23KoAbDi,ref:23GrDiTa,ref:23TaBeDi,ref:24ScHeCh,ref:24ArKaHe}. Importantly, \ce{H2O} emission can be studied in more detail \citep[e.g.][]{ref:sz98,ref:23BaPoCa,ref:24PoSaBa,ref:24TeDiGa,ref:24RoBaOb}, with the first detection of \ce{H2O} in a disc with confirmed planets by \citet{ref:23PeChHe}. The link between inner disc composition and the composition of the planets formed there is still a topic of debate \citep[see][for a review]{ref:21ObBe}; therefore understanding the mechanisms that influence the species available for accretion onto protoplanets in the inner disc is of interest. However, the emission from different discs observed with \textit{JWST} has proven to be diverse, ranging from \ce{H2O}-rich discs and strong silicate features to carbon-dominated discs and weak silicate features \citep{ref:23KaHeAr,ref:23DiGrTa,ref:24HeKaSa}.

Observations of protoplanetary discs using the Atacama Large Millimeter/submillimeter Array (ALMA) demonstrate an equally large variety in outer dust disc structure \cite[e.g.][]{ref:18HuAnDu,ref:18LoPiHe,ref:19LoHeGr}. Some dust discs are very compact \citep[e.g.][]{ref:16AnWiMa,ref:19LoHeGr}, while others are extremely large and contain rings and gaps \citep[e.g. V1094~Sco and Sz~98 in][]{ref:18TeDiAn}, and some even show spiral structures \citep[e.g. IM Lup in][]{ref:18AnHuPe}. In this work, we make the distinction between gaps and cavities, whereby cavities span from the star to the start of the dust disc, while gaps are annular substructures located within the disc. An extensive list of mechanisms exists that can explain the formation of pressure bumps and other substructures \citep[see][for an overview]{ref:23BaIsZh}, one of which is planet formation. We note, however, that the structures within the dust do not have to coincide with that of the gas \citep{ref:21ObGuWa}, or have the same depth, such as for CO and its isotopologues, which are generally less deep \citep{ref:16MaDiBr}.

In the days of \textit{Spitzer}, a range of sample studies pointed to a variety of correlations between the outer dust disc structure and molecular emission from the inner 0.1-10~au of the disc, mainly with \ce{H2O} gas emission. For example, \citet{ref:11CaNa} and \citet{ref:13NaCaPo} examined the ratio of \ce{HCN} to \ce{H2O} as a volatile \ce{C/O} tracer, and found it to increase with increasing disc mass. They attribute this to sequestration of \ce{H2O}-ice in the outer disc due to planetesimal formation, which would be more efficient for more massive discs. Similarly, \citet{ref:20BaPaBo} find \ce{H2O} line fluxes to be stronger for smaller dust disc sizes, which is a potential tracer for efficient pebble drift delivering \ce{H2O} to the inner disc. In these works and the work presented here, we differentiate between the larger dust pebbles that move decoupled from the gas, and smaller micron-sized dust grains. Recently, \citet{ref:23BaPoCa} put this to the test using four T~Tauri discs observed with MIRI/MRS. Their sample, consisting of two extended dust discs (IQ~Tau and CI~Tau) and two compact discs (GK~Tau and HP~Tau), seems to follow this anti-correlation between \ce{H2O} line flux and dust disc size. Additionally, they find a cold reservoir of \ce{H2O} (near sublimation temperatures; 170-400~K) in the smaller discs, indicative of pebble drift. They posit, based on for example \citet{ref:09MePoBl,ref:18KrScBe}, that without pebble drift the colder regions in the disc would quickly be depleted of \ce{H2O}-gas due to the cold-finger effect trapping \ce{H2O} in the form of ice. Drift would replenish this reservoir and produce a cold excess due to sublimation and diffusion at the snowline. One additional important factor is gap formation \citep{ref:17BaPoSa}. Pebble drift can be prevented by substructures like gaps and rings in the outer disc. Sufficiently deep gaps can trap icy pebbles outside the \ce{H2O} snowline, resulting in weak \ce{H2O} emission and a lack of a cold reservoir. All of this indicates that there is observational evidence for the influence of the dust disc structure, like that derived from ALMA, on the ice species that reach the inner disc, and by extension its chemical composition, as has previously been seen with \textit{Spitzer}, and now MIRI/MRS.

From a theoretical point of view, some modelling efforts exist that aim to link the evolution of the dust disc structure to the abundances of gas species in the inner disc. For example, \citet{ref:17BoClMa} and \citet{ref:19BoIl} find that when transport of dust is sufficiently rapid, the abundance of gas species are enhanced within their respective snowlines due to sublimation from incoming ices. Initially, due to the sequence in \ce{H2O}, \ce{CO2}, and \ce{CO} snowlines from closer to farther away from the star, the gaseous \ce{C/O} and \ce{C/H} ratios in the inner disc become sub- and super-solar, respectively. Over time, the \ce{H2O}-gas will be accreted onto the star first, after which \ce{CO2} and \ce{CO} are accreted, resulting in a more carbon-dominated inner disc as the disc evolves further \citep[see also][]{ref:23MaBiPa}. 

\citet{ref:24VlDiTa} model the influence of a cavity on the \ce{H2O} and \ce{CO2} content within the disc. They show that if the cavity extends past the \ce{H2O} snowline, \ce{H2O} emission will be suppressed, and the resulting spectrum is dominant in \ce{CO2} instead. This variation in inner disc emission with cavity size was also observed by \citet{ref:18WoMiTh}.

Alternatively, \citet{ref:21KaPiKr,ref:23KaPiKr} examine the influence of pebble drift in more detail for \ce{H2O}, with the presence of gaps within the disc included, rather than a cavity. They find that the gaseous enhancement, at least for \ce{H2O}, is temporary, lasting until the \ce{H2O} gas has been accreted onto the central star. This means that the phase for which \ce{H2O} is actually enhanced is relatively short-lived (up to a few million years or less), although this might be extended significantly if the gap only partially blocks the influx of dust pebbles \citep{ref:24MaSaBi} or even smaller dust grains \citep{ref:24PiBeWa}. According to \citet{ref:24SeVlDi}, this enhancement may not be directly visible in column densities derived from MRS spectra, due to an increased dust opacity as a result of drifting dust. They find the column density ratio between \ce{CO2 / H2O} to be a more reliable drift tracer. Furthermore, both \citet{ref:23KaPiKr} and \citet{ref:24SeVlDi} note that the radial location of the gap is of influence: gaps closer to the star are more effective at reducing the \ce{H2O}-gas enhancement, since more dust is located outside the gap, while a gap too far from the star blocks only a small part of the \ce{H2O}-ice reservoir and is less effective. However, in some of the aforementioned models \citep{ref:17BoClMa,ref:19BoIl,ref:24MaSaBi,ref:24SeVlDi} gas can move inwards over time, and even diffuse across gaps. This is in line with the expectations of gaps carved by planet formation, which are likely not efficient at blocking gas and the gaps can be leaky to small dust grains \citep{ref:06LuDa,ref:20BeBi,ref:21KaPiKr,ref:23KaPiKr,ref:24SeVlDi}. On the other hand, \citet{ref:24LiBiHe} also model gaps formed due to photoevaporation by X-rays, which could prevent all transport across the gap, even by gas. While the gap itself continues to only block dust transport, the photoevaporative winds blow gas away from the gap \citep[e.g.][]{ref:14AlPaAn,ref:23PaCaEd}. 


Since the modelling efforts demonstrate that the enhancement can be variable in time, whether or not the increase in \ce{H2O}-line flux becomes visible then depends on when gaps are formed and at what stage in the lifetime of the disc it is observed. Even relatively young discs, such as the young Class~II object HL~Tau \citep{ref:15AlBrPe} and Class~I object IRS~63 \citep{ref:20SeScPi}, can have observable, albeit weak, substructures meaning that diversity in inner disc compositions due to the outer disc structure can already start early in the lifetimes of discs. However, the Class~I objects being studied in the eDisks ALMA large programme \citep{ref:23OhToJo} show no deep, detectable substructure, indicating that their formation likely occurs on the cusp of the transition between the Class~I and Class~II phases. Early substructure formation is supported by recent population synthesis modelling work \citep{ref:24DeBiMi}. We do note that disc and stellar age estimations often still have large uncertainties \citep[e.g.][]{ref:23MiKaBi}.

Contrary to the above, Sz~98 is an example of a large dust disc ($\sim$180~au) with multiple gaps, but is found to be more dominant in \ce{H2O} than carbon-bearing species \citep{ref:sz98}, indicating that the story is more complex. Furthermore, drifting dust pebbles will not only contain \ce{H2O}-ice; other ices should also be drifting inwards along with the dust, and could show some enhancement akin to the findings of \citet{ref:19BoIl}. Due to the sequence in \ce{H2O}, \ce{CO2}, and \ce{CO} snowlines and theorised influence of dust drift on the inner disc composition, it naturally follows that both the presence and radial location of gaps may not only influence the presence of \ce{H2O} in the inner disc, but also that of other oxygen- and carbon-bearing species. However, this relies on ices being released into the gas at their respective snowlines. This notion of sequential snowlines has been pointed out to not be entirely realistic. Lab works show that other volatiles can be trapped in \ce{H2O} ice \citep[e.g.][]{ref:04CoAnCh}, and \ce{H2O} ice in turn can be trapped within refractories, such as silicates \citep{ref:24PoJaMu}. As a result, the gaseous volatile ratios become much less dependent on the radial location in the disc, and instead several snowlines exist per species depending on the layering within the ice. For example, many volatiles are released along with \ce{H2O} at the \ce{H2O} snowline \citep[e.g.][]{ref:04ViCoDe,ref:09ViDiDo,ref:24LiKiGa}, and active chemistry occurs in the cold outer disc \citep{ref:18EiWaDi}. This can transform CO into less volatile species, such as \ce{CH3OH} and \ce{CO2} \citep[e.g.][]{ref:16YuWiDo,ref:18BoWaDi}.

Furthermore, radial transport of dust is not the only possible explanation for differences in emitting strength of species between discs. An increase in emission strength from all species can also be achieved by (1) increasing the ratio between the gas and dust, or (2) more transparent dust due to grain growth or dust settling, where both scenarios result in emission from a larger column of gas \citep[e.g.][]{ref:09MePoBl,ref:15AnKaRi,ref:18WoMiTh}. However, due to the variation in where the majority of different gas species are located spatially, changes in the dust opacity can influence the emission of certain molecules more strongly than others. For example, the relative strength of \ce{HCN} compared to \ce{H2O} increases with a decrease in dust opacity \citep{ref:23AnKaWa}.

In this work, we explore the variation between the MIRI/MRS spectra of a sample of discs, most of which are part of the MIRI mid-INfrared Disk Survey \citep[MINDS][]{ref:24HeKaSa,ref:23KaHeAr}. In order to test whether the presence of a gap close to the star reduces the flux of \ce{H2O} and increases that of carbon-bearing species in the inner disc, we compare the emission from commonly detected species \ce{H2O}, \ce{HCN}, \ce{C2H2}, and \ce{CO2} to the dust disc structure as found in ALMA data. To this end, we reanalyse ALMA continuum data to find the outer dust disc structures in a consistent manner. 

The work presented here is structured as follows. First, the sample used, data reduction method, and lines used in this study are described in Sect.~\ref{sec:methods}. The molecular emission in the sample is presented in Sect.~\ref{sec:results}. We compare these to the dust disc structure and put these in context of disc chemistry and planet formation in Sect.~\ref{sec:discussion}. The summary of the findings can be found in Sect.~\ref{sec:conclusion}.

\section{Methods} \label{sec:methods}

\subsection{Sample}

\begin{table*}[t]
\caption{Sample of T Tauri discs observed with \textit{JWST} included in this work, and assumed properties, organised by increasing stellar mass.}
\label{tab:sample}
\centering
\resizebox{1\linewidth}{!}{%
\begin{tabular}{lcccccccccccc}
\hline \hline
\multicolumn{1}{l}{ID} & \multicolumn{1}{c}{\begin{tabular}[c]{@{}c@{}}$M_{\star}$\\ {[}$M_{\odot}${]}\end{tabular}}  & \multicolumn{1}{c}{\begin{tabular}[c]{@{}c@{}}$L_{\star}$\\ {[}$L_{\odot}${]}\end{tabular}}  &\multicolumn{1}{c}{\begin{tabular}[c]{@{}c@{}}$\log(L_\text{acc})$\\ {[}$L_{\odot}${]}\end{tabular}} & \multicolumn{1}{c}{\begin{tabular}[c]{@{}c@{}}Distance\\ {[}pc{]}\end{tabular}} & \multicolumn{1}{c}{\begin{tabular}[c]{@{}c@{}}$R_\text{dust}$\\ {[}au{]}\end{tabular}} & \multicolumn{1}{c}{\begin{tabular}[c]{@{}c@{}}$R_\text{gap}$\\ {[}au{]}\end{tabular}} & PID & \multicolumn{1}{c}{\begin{tabular}[c]{@{}c@{}}Date observed\\ {[}d/m/y{]}\end{tabular}} & \multicolumn{1}{c}{\begin{tabular}[c]{@{}c@{}}$\sigma$ \@ $\sim$15.9~$\mu$m\\ {[}mJy{]} \end{tabular}} & \multicolumn{1}{c}{\begin{tabular}[c]{@{}c@{}}Age\\ {[}Myr{]}\end{tabular}}  \\ \hline
Full discs \\ \hline
GW Lup   &  0.46$^{A1}$  & 0.33$^{L1}$ & -2.17$^{L1}$        & 155                                                                                      & 102.3                                                                                  & 47.5 &   1282                                                                               & 8/8/2022          &     0.7    & 0.8-5.0$^{A1}$                    \\
IQ Tau   & 0.50$^{L2}$  & 0.22$^{L3}$ &  -1.40$^{L5}$       & 132                                                                                      & 103.0                  & 39.3 &   1640                                           & 27/2/2023   &  1.3 & 2.2-8.3$^{L2}$   \\
BP Tau    &  0.52$^{L2}$  & 0.4$^{L2}$ &  -1.17$^{L2}$        & 127                                                                                      & 40.6                                                                                    & -  &  1282                                           & 19/2/2023    &  1.5 & 1.0-3.4$^{L2}$  \\
Sz 98   & 0.74$^{L1}$  & 1.52$^{L1}$ & -0.72$^{L1}$       & 156                                                                                      & 147.3                                                                                   & 8.6  & 1282                                                                             & 8/8/2022      &     1.5     & 1.7-5.6$^{A2}$             \\
CI Tau    & 0.89$^{L2}$  & 0.83$^{L3}$ &     -0.70$^{L4}$   &     160                                                         &   172.5 & 15.2 &     1640                                 &  27/2/2023  &  1.5 & 1.4-4.5$^{L2}$ \\
V1094 Sco    & 0.92$^{M1}$ & 1.17$^{L1}$ &  -1.02$^{L1}$    & 155                                                                                  & 239.3                                                                                  & 15.6  & 1282                                                                             & 9/8/2022        &       0.3        & 6.0-18$^{A2}$                                            \\
DR Tau     & 0.93$^{L2}$  & 0.63$^{L2}$ & -0.24$^{L2}$        & 193                                                                                      & 54.6                                                                                   & 37.7   &    
1282                                         & 3/3/2023          &         8.4     & 1.8-5.9$^{L2}$                                  \\
DL Tau      &  0.98$^{L2}$  & 0.65$^{L2}$  &  -0.47$^{L2}$    & 160                                                                                      & 147.2                                                                                   & 13.8     &    1282                                                & 27/2/2023        &    1.2       & 1.9-6.3$^{L2}$                     \\ \hline
Transition discs  &  &  &  & & & $(R_{\text{ring}})$  \\ \hline
PDS 70   &  0.76$^{A3}$  &  0.35$^{L6}$  & -2.81$^{L7}$ &   114    &                                                                                      100             & 70  &   1282 &   1/8/2022    &     0.3     &     4.4-6.4$^{A3}$         \\
SY Cha   & 0.78$^{L8}$  &  0.47$^{L8}$ & -2.24$^{L8}$  &   180   &     135 &                                                                                  100 &   1282                                                                               &  10/8/2022                                                              &  0.7  & 3$^{A4}$    \\ \hline
\end{tabular}%
}

\vspace{1ex}

     {\raggedright \textbf{Notes.} Masses from $^{M1}$\citet{ref:18TeDiAn}. $L_\star$ and $L_{\text{acc}}$ from $^{L1}$\citet{ref:17AlMaNa}, $^{L2}$\citet{ref:19LoHeGr}, $^{L3}$\citet{ref:14HeHi}, $^{L4}$\citet{ref:20DoBoAl}, $^{L5}$\citet{ref:22GaAnBi}, $^{L6}$\citet{ref:16PeMa}, $^{L7}$\citet{ref:22SkAu}, and $^{L8}$\citet{ref:16MaFeHe} scaled to the updated \textit{Gaia} distance where relevant. All distances from the \textit{Gaia} DR2 parallax \citep{ref:gaia1,ref:gaia2}. Outer dust and gap radii estimated from ALMA data (see App. \ref{app:ALMA}). Ages are from $^{A1}$\citet{ref:18AnHuPe}, $^{A2}$\citet{ref:19MaDoFr}, $^{A3}$\citet{ref:18MuKeHe}, and $^{A4}$\citet{ref:23OrMoMu}. \par}
     {\raggedright Ages are generally very uncertain and may be based on different estimation methods. Especially for Sz~98 and V1094~Sco, we found large discrepancies in the literature, and we assume they are of the order of millions of years old, similarly to the other discs. \par}


\end{table*}

The sample presented here is a selection of discs around T~Tauri stars, from two \textit{JWST} programmes. The majority of the sources come from the MINDS programme, which is one of the Guaranteed Time Observation programmes of \textit{JWST} (PID: 1282, PI: T. Henning). The total sample consists of 52 targets, out of which 33 are discs around T~Tauri stars \citep{ref:23KaHeAr,ref:24HeKaSa}. The second is PID 1640 (PI: A. Banzatti), which consists of eight different targets. Out of these 41 T~Tauri discs, we excluded discs known to be in binary systems, discs with confirmed spiral features in millimetre emission, and highly inclined discs ($i>$70$\degree$). Furthermore, it has already been demonstrated that discs of very low-mass objects show significantly different emission features, perhaps due to a difference in evolution timescale as compared to objects of higher mass \citep[e.g.][]{ref:23MaBiPa}, resulting in extremely hydrocarbon-rich discs \citep[e.g.][Morales-Caleron, in prep.]{ref:23TaBeDi,ref:24ArKaHe,ref:24KaKaJa}. For gap formation to then influence the composition of the inner disc, gaps must form even earlier in the disc's lifetime. Furthermore, due to the lower temperatures in these discs, gaps must be located much closer to the central star in order to still block oxygen-rich ices. To ensure our sample is homogeneous, the objects are selected to have at least M$_\star \geq$ 0.45M$_{\odot}$.

Since our aim is to study the influence of gaps, only discs with confirmed gaps in millimetre dust are included, based on previously published results \citep{ref:18ClTaJu,ref:18TeDiAn,ref:21RoIlFa,ref:22JeBoTa,ref:22JeTaCl,ref:23ZhKaLo}. After carefully selecting the sample based on the aforementioned criteria, eight full discs from the two programmes remain and are listed in Table~\ref{tab:sample}. Six of these are from the MINDS programme, along with two discs out of the eight targets from PID~1640, for which the \ce{H2O} data are presented in \citet{ref:23BaPoCa}. We further compare these cases to discs with extremely wide gaps of several tens of astronomical units, where transport may be most limited: two discs from MINDS (PID~1282), PDS~70 and SY~Cha. Despite the large gaps, both PDS~70 and SY~Cha have \ce{H2O} emission in the inner disc \citep{ref:23PeChHe,ref:24ScHeCh}, and signs for transport across their large gaps \citep{ref:23OrMoMu,ref:24JaWaKa}. For these reasons, they are interesting additions to the sample presented here. The associated PID and date of observation per target and other properties are summarised in Table~\ref{tab:sample}.


\subsection{Data reduction}
For consistency, all data were reduced using version 1.15.1 of the standard \textit{JWST} pipeline \citep{ref:23BuEiDe}, and pmap 1252. While targets observed at the start of the MINDS programme make use of target acquisition, a large part of the sample taken later in the year did not. As such, the data cannot consistently be processed based on the methods presented by \citet{ref:23GaArSl} as was done in \citet{ref:23GrDiTa} and \citet{ref:sz98}, or using asteroids as in \citet{ref:23BaPoCa} and \citet{ref:24PoSaBa}. To ensure that differences in the data reduction methods do not affect the relative fluxes between targets, we used the standard pipeline set-up, with extended fringe flats and residual fringe correction both on the detector and spectrum level, to reduce the data. The spectra are extracted from the cube using a growing circular aperture of 2$\times$FWHM (`full-width at half maximum') centred on the source, and an annulus to estimate the background, as is the standard procedure in the current pipeline. For the majority of the sample this is currently the best method, after the many updates since \citet{ref:23GrDiTa} and \citet{ref:sz98}. The resulting fluxes may therefore differ slightly from the aforementioned works, but it ensures that we are consistent within the sample shown here. When comparing the difference in flux between the spectrum of GW~Lup here and in \citet{ref:23GrDiTa}, this ranges from less than a percent at the shorter wavelengths, to under 3\% in the longest. This discrepancy is smaller when comparing \citet{ref:sz98} to the spectrum of Sz~98 shown here, which ranges from less than a percent at the shorter wavelengths, to up to a percent in the longest. The slightly larger discrepancy in the longer wavelengths makes sense, since it has since been discovered that a drop in response has occurred since the start of operation. This is now accounted for in the current pipeline version, and is most significant at the longer wavelengths \citep{ref:24LaArGo}.

While the data reduction methods are consistent between targets, the set-up of the observations is not. However, the noise ($\sigma$, measured as standard deviation on the featureless spectrum) is very similar between targets overall, as is shown in Table~\ref{tab:sample}. The noise is measured on the spectra around 15.9~$\mu$m directly after subtracting the rough slab fits (see Sect.~\ref{sec:line_int}). These slab fits are not perfect and may leave small residuals from molecular features, which would then slightly affect the estimated noise. It is therefore a rough estimate. DR~Tau is the only outlier in this sample, but this could be due to the fact that finding a line-free region in its extremely line-rich spectrum is difficult, and therefore larger residuals from line emission are included in the measurement (see \citealt{ref:24TeDiGa}). The total observing time of DR~Tau is similar to that of DL~Tau and BP~Tau, whilst being brighter than both. While the data may be slightly noisier due to the few number of groups per integration than a more ideal set up, the true noise level is likely similar to that of these two objects. Regardless, we proceed to use the measured value when determining the error bars.

The continua in the spectra, which we define as the SED including dust features, were also subtracted in a consistent manner. In order to do so, this has been automatised in the manner described in \citet{ref:24TeDiGr}, which is based on baseline estimations using \texttt{PyBaselines} \citep{ref:22Er}. First, a Savitzky-Golay filter was iteratively applied to filter out $>2\sigma$ emission lines, until no more outliers were identified. Negative spikes were removed afterwards. These downward spikes were masked in the original spectrum. Subsequently, the `iterative reweighted spline quantile regression' (IRSQR) method from \texttt{PyBaselines} was used to estimate the continuum. For the objects included here this works well, since most of the gas features are superimposed directly on top of the dust continuum.

\subsection{Line fluxes}
\label{sec:line_int}
We used the line fluxes of other molecules, most notably \ce{C2H2}, as volatiles C/O tracers. The ranges over which the fluxes of \ce{C2H2}, \ce{HCN}, and \ce{CO2} were measured are presented in Table~\ref{tab:carb_flux}, and encompass the emission from the \textit{Q}-branches. The flux of the continuum-subtracted spectrum is integrated in these wavelengths to determine the line flux. These ranges have been adjusted slightly compared to \citet{ref:11SaPoBl}, to limit some blending with other species.

\begin{table}[t]
\caption{Wavelength ranges used to measure the integrated flux of common carbon-bearing species.}
\label{tab:carb_flux}
\centering
\begin{tabular}{lc}
\hline \hline
                          & \begin{tabular}[c]{@{}c@{}}Wavelength region\\ {[}$\mu$m{]} \end{tabular} \\ \hline
\ce{CO2} &  14.936--15.014                                                           \\
\ce{HCN} & 13.837-–14.075                                                             \\
\ce{C2H2} & 13.61--13.751                                                             \\ \hline
\end{tabular}

\end{table}

For \ce{H2O}, we focussed on the rotational \ce{H2O} lines, since the ro-vibrational lines tend to deviate more from local thermodynamic equilibrium (LTE) excitation \citep{ref:09MePoBl}. \citet{ref:23BaPoCa} use the relative strength of the total flux of isolated warm lines (upper energies, $E_{\rm{up}}$, of $6000 \leq E_{\rm{up}} < 8000$~K) compared to the total flux of isolated cold lines ($E_{\rm{up}} < 4000$~K) as an indicator of efficient drift. The improved resolution and sensitivity of MIRI/MRS now make it possible to identify and characterise individual transitions, rather than just line blends, and we exploited this further in this work. Following \citet{ref:23BaPoCa}, we included a range of isolated warm lines (upper energies, $E_{\rm{up}}$, of $6000 \leq E_{\rm{up}} < 8000$~K) and isolated cold lines ($E_{\rm{up}} < 4000$~K). For all lines, they are either isolated, or have a significantly higher $A_{\rm{ul}}$ than their neighbours, and we can assume they dominate the flux in that part of the spectrum. In order to make sure that the upper state degeneracy ($g_{up}$) did not cause neighbouring lines to be brighter than expected from their $A_{\rm{ul}}$, LTE spectra were generated to check that individual transitions were indeed significantly brighter than any neighbouring lines for a range of temperatures and column densities. While a wide range of these lines exist, we only included the few that are most widely detected in our sample. This limits the list to those presented in Fig.~\ref{fig:lines_used} and Table~\ref{tab:lines}, with a total of 13 individual transitions. The line list is smaller compared to \citet{ref:23BaPoCa} due to our imposed requirement that the lines must be both isolated and detectable in our sample, which includes sources with relatively weak H$_\text{2}$O emission. The trends presented here may therefore be slightly more affected by noise.

As is demonstrated for \ce{CO2} in \citet{ref:23GrDiTa}, the line flux of a bright feature is also affected by emitting area, rather than just column density. Therefore, it is the weaker features and ratios between individual lines that will allow us to unambiguously compare column densities. The emission from a certain molecule can be strong in a disc, but this could simply be due to a large emitting area, which also scales the flux, rather than a high column density (see also Arabhavi et al., in prep.). Lines of the same upper level, but different lower level and Einstein coefficient ($A_{\rm{ul}}$), originate from the same region in the disc and information about the column density can be inferred from their flux ratio. If both lines are optically thin, the flux ratio will simply be equal to the ratio of the Einstein coefficients ($A_{\rm{ul}1}/A_{\rm{ul}2}$). However, as soon as one of the lines saturates, the ratio will deviate from this quantity, and scale with column density until both lines become optically thick.

This is done for \ce{H2O} in \citet{ref:sz98}. Here, we identified a different pair of relatively isolated cold lines (at 13.503 and 22.375~$\mu$m, transitions 11$_{\text{ }7\text{ }4}$--10$_{\text{ }4\text{ }7}$ and 11$_{\text{ }7\text{ }4}$--10$_{\text{ }6\text{ }5}$), which are indicated by the triangles in Fig.~\ref{fig:lines_used} and documented in Table~\ref{tab:lines}. This way, the difference between colder reservoirs of \ce{H2O} in the discs could be examined more directly than just by using line fluxes. Similarly to \citet{ref:sz98}, the line ratios are compared to the ratio in slab models of varying temperatures to find what column density the line ratio corresponds to. Based on the multiple temperature component models in \citet[][opacities from private communication]{ref:24TeDiGa}, we can conclude that, at least for DR~Tau, this line pair traces the intermediate temperature component ($\sim$400~K). Furthermore, based on these same opacities derived from the component models in \citet{ref:24TeDiGa}, the 11$_{\text{ }7\text{ }4}$--10$_{\text{ }4\text{ }7}$ transition is likely not quite optically thick, while the 11$_{\text{ }7\text{ }4}$--10$_{\text{ }6\text{ }5}$ is. Since the lines tracing the coldest component in DR~Tau are not optically thick, it is not possible to compare column densities in this region. We note that during the reviewing process of this work, the same column density line pair was identified independently in \citet{ref:24BaSaPo}. The detection of the isotopologue H$_2^{18}$O would fully constrain the column density.

To minimise the effect of blending with emission of other molecules in the \ce{HCN}, \ce{CO2}, and \ce{C2H2} line fluxes, slab fits of \ce{H2O}, \ce{CO2}, \ce{C2H2}, \ce{HCN}, and \ce{OH} were subtracted prior to integrating the flux in this area. The error in integrated flux is based on the $\sigma$ values given in Table~\ref{tab:sample}. The fits were performed in the same way as those presented in other MINDS works \citep[e.g.][]{ref:23GrDiTa,ref:23PeChHe,ref:sz98}, but we performed the fits again in order to make sure the percent-level flux differences from the change in data reduction do not influence the residuals. We refer the reader to these works for further details. This becomes especially important for \ce{HCN}, which is blended with \ce{C2H2} and \ce{H2O}, as well as some \ce{OH}. We note that the slab fits were used solely for the purpose of removing other molecular spectral features; we do not derive any excitation properties from the slab fits.

Finally, previous works have found the \ce{H2O} emission strength to correlate with $\log{(L_\text{acc})}$, with a slope of $\sim$0.6. We divide by this factor to remove any influence of enhanced emission due to an increase in $L_\text{acc}$ \citep{ref:20BaPaBo,ref:23BaPoCa}. Furthermore, when not considering line ratios, the fluxes are rescaled to a consistent distance of 140~pc.

\subsection{Correlation coefficient and p value}
\label{sec:corr}
In order to determine how significant the negative correlation between \ce{H2O} line flux and outer dust disc radius \citep{ref:20BaPaBo,ref:23BaPoCa} is in the sample presented here, we use the Pearson correlation coefficient (PCC) and corresponding p value from \texttt{scipy} \citep{scipy}. We consider a PCC of $\geq 0.8$ to be a very strong correlation, and a PCC of $\geq 0.4$ to be moderately correlated. When calculating the PCC, we do not include SY~Cha and PDS~70, as these are outliers in terms of disc structure. The p value is defined as the likelihood that any other randomly distributed sample would have the same, or an even stronger PCC, where a p value of less than 0.05 indicates that the result is statistically significant. However, we note that these quantities must be used with some caution, as the sample size is too small to give conclusive results.

\subsection{DR Tau as a spectral template}
\label{sec:spec_temp}
The MIRI/MRS spectrum of DR~Tau has been studied more extensively in \citet{ref:24TeDiGa,ref:24TeDiGr}, and multiple temperature component fits of \ce{H2O} demonstrate the presence of cold \ce{H2O} emission out to $\sim$6-8~au. DR~Tau is overall a very \ce{H2O}-rich disc, with weaker detections of \ce{C2H2}, \ce{CO2}, and \ce{HCN}. More detailed power law fits with $T\propto R^{-0.5}$ (similar to \citealt{ref:24RoBaOb}) demonstrate that there is likely no extreme excess in the cold \ce{H2O} abundance, and DR~Tau is in a more regular drift regime, close to GK~Tau, when placed on Fig.~9 in \citet{ref:24BaSaPo} (Temmink et al. in prep.). As it is well studied and its emission relatively well characterised, we used DR~Tau as a template to compare to the emission of the other discs. Here, it must be kept in mind that DR~Tau is an \ce{H2O}-rich source is a source rich in \ce{H2O} lines, with some signs of \ce{H2O}-enhancement at the snowline, but not particularly large. This comparison helped us identify which discs show clear excesses or depletion in low energy \ce{H2O}, focussing on the 23.8 to 24~$\mu$m \ce{H2O} quadruplet most dominantly affected by the coldest \ce{H2O} component \citep{ref:24TeDiGa}. Furthermore, discs that display relative strong emission from \ce{HCN}, \ce{C2H2}, and \ce{CO2}, could also be clearly identified.

In order to use DR~Tau as a template, we followed the approach in \citet{ref:23BaPoCa}, and rescaled all the spectra to a consistent distance of 140~pc, and scaled the DR~Tau spectrum to the flux levels of the other discs by using the relative strength of the $6000 \leq E_{\rm{up}} < 8000$~K \ce{H2O} lines (H$_2$O$_{\text{warm, disc ID}}$/H$_2$O$_{\text{warm,DR Tau}}$).

\begin{figure*}[t]
    \centering
    \includegraphics[width=\textwidth]{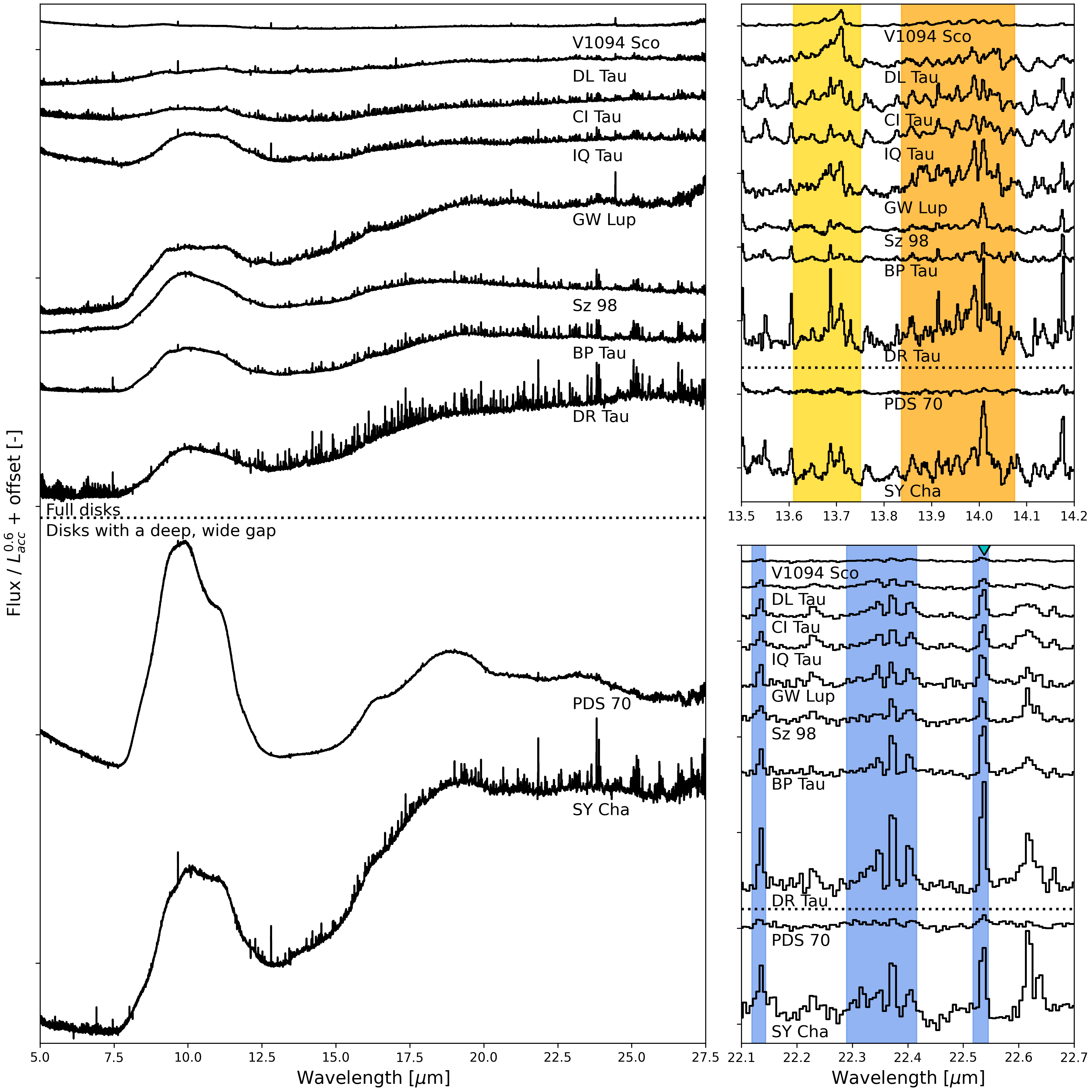}
    \caption{Spectra of all discs considered in this study. The fluxes have been rescaled with respect to distance and $L_{\text{acc}}^{0.6}$, the full discs are arranged from largest \ce{H2O}$_\text{cold}$/$L_\text{acc}^{0.6}$ (bottom), to smallest (top). The right column shows sections of the spectra with \ce{C2H2} (yellow shading) and \ce{HCN} (orange shading) emission (top), and \ce{H2O} (blue shading) emission (bottom). The triangle in the bottom right panel indicates the line in this wavelength region used to calculate the cold \ce{H2O} flux. The shaded regions of \ce{C2H2} and \ce{HCN} are also the regions over which the integrated fluxes are calculated.}
    \label{fig:all_specs}
\end{figure*}

\subsection{ALMA}
The outer dust disc and gap radii are taken from the visibility fits of ALMA continuum Band 6 data (see Table~\ref{tab:sample}). For many of the discs used in this sample, an analysis of the ALMA data already exists. However, this is scattered in various works using various methods. Therefore, to ensure all radii and substructures are retrieved in a consistent manner for all discs, the analysis has been redone here using the highest resolution archival data available. By fitting the visibility plane with sets of Gaussians (following the approach in \citet{ref:16ZhBeBl}), less obvious substructures are revealed. A detailed discussion of the datasets used, fitting method, and results of the fits can be found in App.~\ref{app:ALMA}. For the outer dust disc radius, the value that encompasses 90\% of the millimetre emission is assumed \citep[e.g.][]{ref:23MiKaBi}. Gaps and rings are identified as local minima and maxima in the radial profile in the visibility plane, though we note that several plateaus (a flattening in the radial profile not surrounded by local minima) are also identified. The radial location of the gap is then defined as the distance from the star to the centre a local minimum. We only document the distance of the gap located closest to the star in Table~\ref{tab:sample} under $R_\text{gap}$, since the innermost gap has been shown to influence the inner disc composition the most \citep{ref:23KaPiKr}. For SY~Cha and PDS~70, where the gaps span several tens of astronomical units, the distance to the ring is given ($R_\text{ring}$) instead. For all identified gaps we refer the reader to App.~\ref{app:ALMA}.

In general, the substructures and radii agree with the values found in previous works \citep[e.g.][]{ref:18TeDiAn,ref:22JeBoTa,ref:22JeTaCl,ref:23ZhKaLo}. We were able to identify some new substructures within 20~au from the star, reporting a new gap at 16~au for V1094~Sco, and a small inner cavity in BP~Tau.

\begin{figure*}[t]
    \centering
    \includegraphics[width=\textwidth]{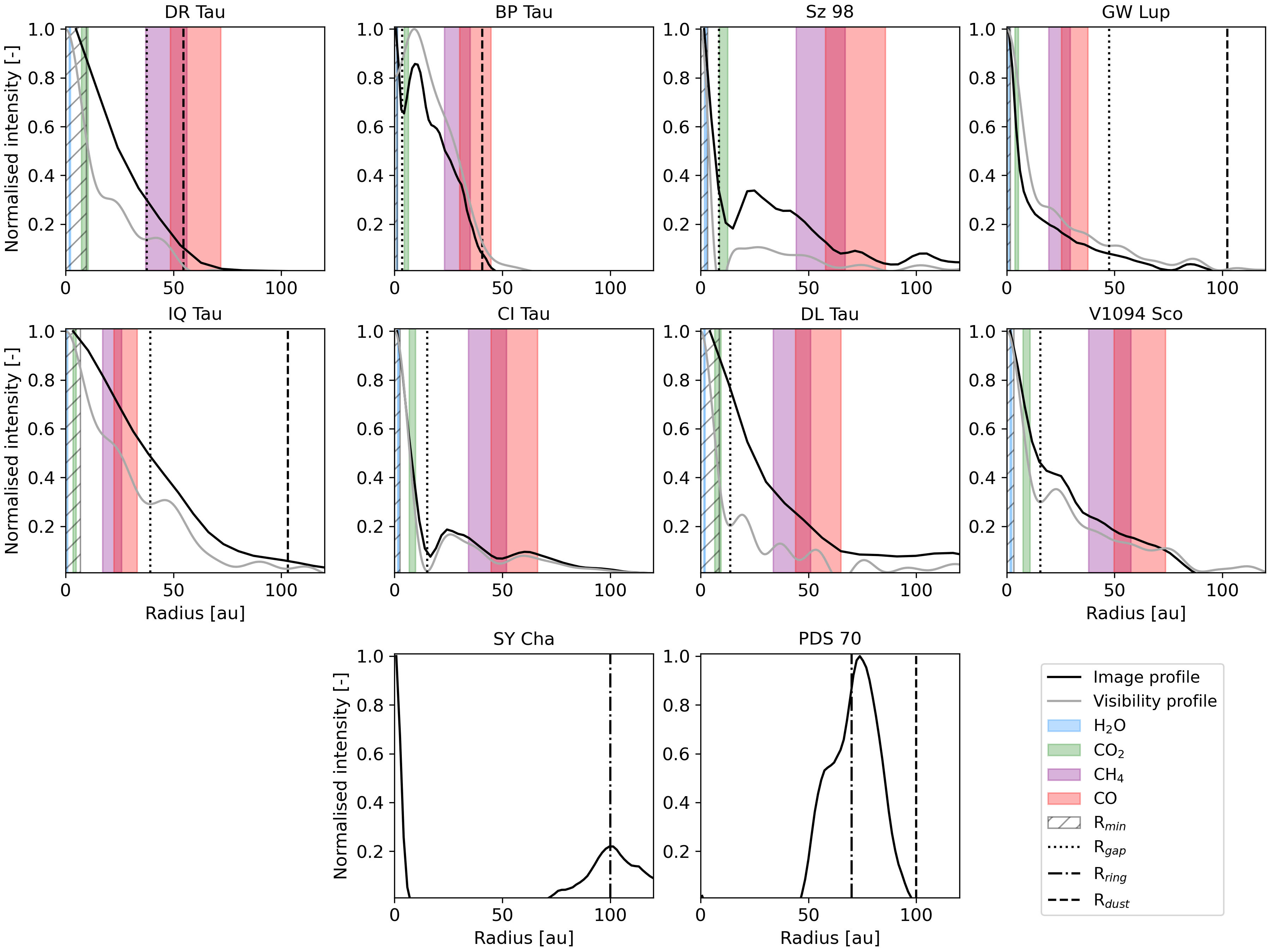}
    \caption{Radial profiles of the disc sample based on the analysis in the image (black line) and visibility plane (grey line). The snowline estimates of \ce{H2O}, \ce{CO2}, \ce{CH4}, and \ce{CO} are indicated by the coloured areas. The minimum distance from the star that can be imaged, based on half the beam radius, is indicated by the hatched area. The $R_\text{gap}$, or the location of the innermost detected local minimum, is indicated by the vertical dotted line. The discs are organised from strongest \ce{H2O}/$L_\text{acc}^{0.6}$ (top left) to weakest (bottom right), with the two (pre-)transition discs in the bottom row.}
    \label{fig:radial_profiles}
\end{figure*}

\begin{table}[t]
\caption{Freeze-out temperatures of selected species used to estimate snowline locations.}
\label{tab:Tcond}
\centering
\begin{tabular}{lc}
\hline \hline
                          & \begin{tabular}[c]{@{}c@{}}$T_{\text{ice}}$\\ {[}K{]}\end{tabular} \\ \hline
\ce{H2O} & 128--155$^{a}$                                                            \\
\ce{CO2} & 60--72$^{a}$                                                              \\
\ce{CH4} & 26--32$^{b}$                                                              \\
\ce{CO}  & 23--28$^{a}$                                                            \\ \hline
\end{tabular}

\vspace{1ex}

     {\raggedright \textbf{Notes.} $^{a}$\citet{ref:14MaMuBu}, $^{b}$\citet{ref:14LuSaSa} \par}

\end{table}

\subsection{Snowline estimates}
Along with the radial profiles, we give rough estimates of the snowline locations of \ce{H2O}, \ce{CO2}, \ce{CH4}, and \ce{CO}. These are calculated as in \citet{ref:18LoPiHe}, using
\begin{equation}
	T(r) = T_{\star} \left( \frac{R_{\star}}{r} \right)^{1/2} \phi_{\text{inc}}^{1/4} \text{ ,}
\end{equation}
from \citet{ref:87KeHa}, assuming a flaring angle of $\phi_{\text{inc}}=0.05$ \citep{ref:04DuDo} and that $T_{\star} = \left(\frac{L_{\star}+L_\text{acc}}{4\pi R_{\star}^{2} \sigma}\right)^{1/4}$, which is most accurate for the midplane temperature. $T(r)$ is the temperature of the disc midplane as a function of radius, $T_{\star}$ the stellar temperature, $R_{\star}$ the stellar radius, $r$ the radial distance measured from the central star, $L_{\star}$ the stellar luminosity, and $\sigma$ the Stefan-Boltzmann constant. Combining these assumptions, we get the following for the snowline locations as a function of the freeze-out temperature ($T_{\text{ice}}$):
\begin{equation}
	r_\text{ice} (T_{\text{ice}}) = \left(\frac{4\pi \sigma T_{\text{ice}}^4}{\phi_{\text{inc}} (L_{\star}+L_\text{acc})}\right)^{-1/2} \text{.}
\end{equation}
Since $r_\text{ice}$ scales with the square root of $\phi_{\text{inc}}$, a change in this quantity does not significantly impact the snowline location.

The assumed ranges of freeze-out temperatures of the species considered can be found in Table~\ref{tab:Tcond}. We note that the snowlines for SY~Cha and PDS~70 are not included, since simple power laws no longer hold for these discs where the inner disc is very small and/or weak. The snowline of \ce{H2O} may fall within the inner disc still, but the other snowlines cannot reliably be calculated, which in our presented resolutions are more informative.

\section{Results} \label{sec:results}

\subsection{Spectra of disc sample}

In Fig.~\ref{fig:all_specs}, we present the MIRI/MRS spectra of the discs considered in this sample, scaled to a distance of 140~pc and divided by $L_\text{acc}^{0.6}$. The discs are organised from strongest to weakest \ce{H2O} emission from bottom to top. In the two panels on the right, some zoomed-in cuts of the spectra are shown to highlight the differences in emission of the carbon species (\ce{C2H2} and \ce{HCN}, top) and \ce{H2O} (bottom). The spectra of the discs with wide gaps (PDS~70 and SY~Cha) are separated from the other discs by a dashed horizontal line. The spectra demonstrate the variety in molecular emission in this sample, with DR~Tau, BP~Tau and Sz~98 showing weaker \ce{C2H2} and \ce{HCN} emission compared to \ce{H2O}. On the other hand, V1094~Sco and DL~Tau are relatively weak in \ce{H2O}, but quite strong in \ce{C2H2} and \ce{HCN}. The spectra of both these discs will be discussed in more detail in Tabone et al. (in prep.). Indeed, the dichotomy presented by models where discs are either rich in \ce{H2O} or carbon seems to tentatively appear in these plots, even for $M_\star \geq 0.45 M_\odot$.

\subsection{Radial profiles and snowlines}
\label{sec:radial_profiles}

The radial profiles derived from the analysis of the ALMA data in the image and visibility plane are shown in Fig.~\ref{fig:radial_profiles}. The discs are organised by relatively strong to weak cold \ce{H2O} line flux from left to right, and top to bottom, with the two discs containing very wide gaps (SY~Cha and PDS~70) in the bottom row. The difference in the ALMA intensity and visibilty profiles arise due to the different resolutions, as the visibility profiles have been created at a higher resolution. Inferring a model profile by convolving the visibility fit with a Gaussian beam, tailored after the ALMA beam, results in a similar profile as that of the ALMA image plane. 

First, we note that the discs with the strongest cold \ce{H2O} emission do have gaps within 20~au from the star. In fact, Sz~98 shows the deepest substructure of the full discs, and is one of the more \ce{H2O}-rich sources, as we discuss in Sect.~\ref{sec:res_flux}. Second, in some discs the estimated \ce{CO} and \ce{CH4} snowlines are clearly present outside of the detected substructures. This is the case for DR~Tau, BP~Tau, Sz~98, CI~Tau, DL~Tau, and V1094~Sco; although for DL~Tau, and V1094~Sco these appear to be relatively shallow substructures. The reverse is true for IQ~Tau and GW~Lup, which have their \ce{CO} and \ce{CH4} snowlines within $R_\text{gap}$. While the substructure in BP~Tau appears as a close-in gap in the image plane, it is best fit with a cavity in the visibility plane. The snowline estimates based on a power law may therefore be less reliable. The same is true for SY~Cha and PDS~70, where we forgo the snowline estimation altogether. 


\begin{figure}[t]
    \centering
    \includegraphics[width=0.6\columnwidth]{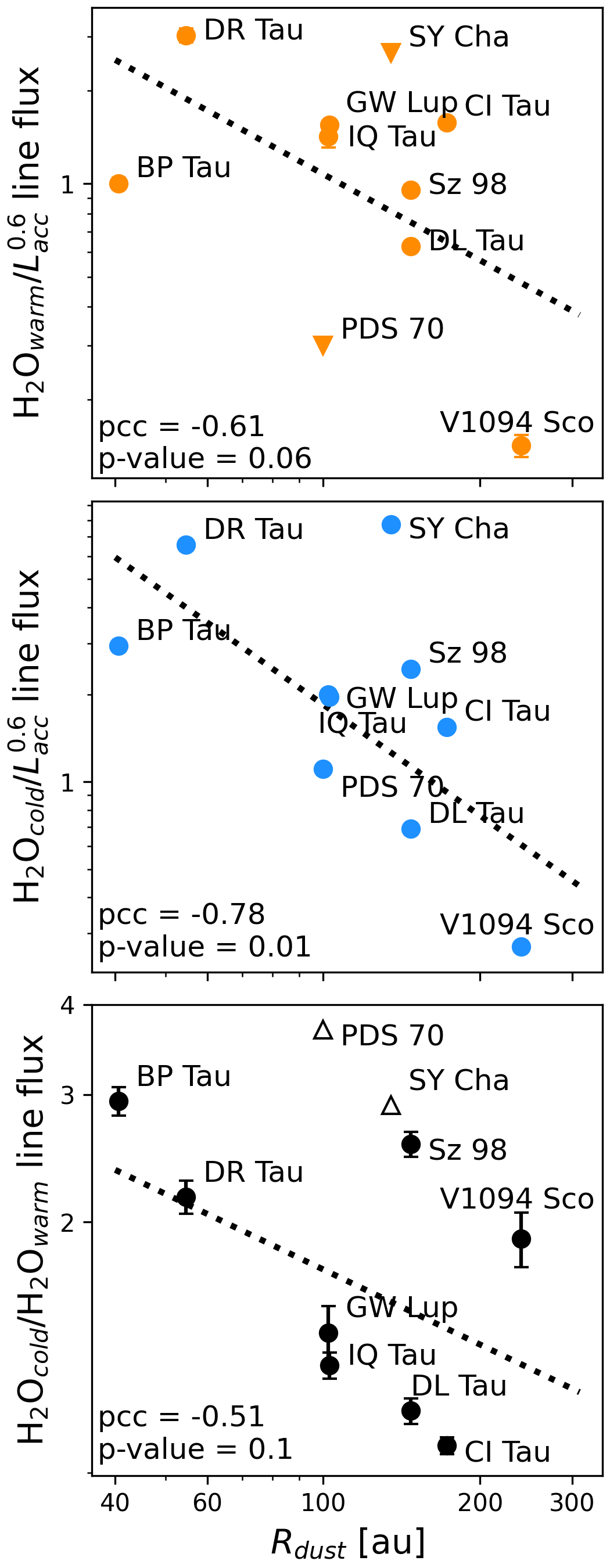}
    \caption{Average of integrated line fluxes of warm (top, orange) and cold (middle, blue) \ce{H2O} gas, and their ratio (bottom, black), plotted against the outer dust disc radius. The PCCs and p values for these log-log scatter plots are given in the panels, and the corresponding trend is indicated by the dotted black line. Upper and lower limits are indicated by triangles.}
    \label{fig:water_radius_corr}
\end{figure}

\subsection{Molecular line fluxes}
\label{sec:res_flux}

\subsubsection{H$_2$O}

\begin{figure}[t]
    \centering
    \includegraphics[width=\columnwidth]{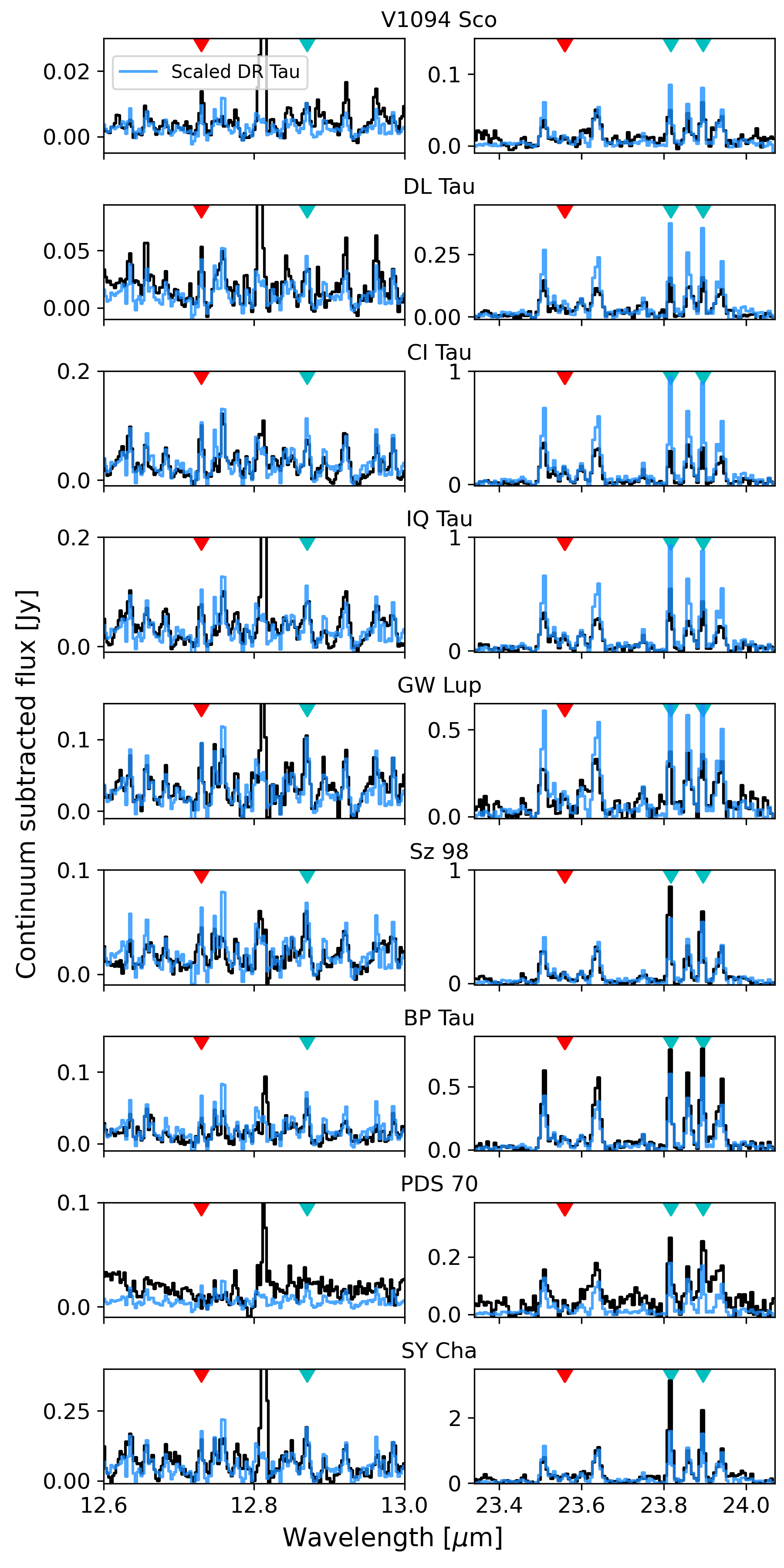}
    \caption{Comparison of warm (indicated by the red triangle) and cold (indicated by the blue triangle) \ce{H2O} lines between DR~Tau rescaled to the line strength of the other discs (blue line), and the emission from the discs themselves (black line).}
    \label{fig:water_excess}
\end{figure}

The line fluxes of all the \ce{H2O} lines mentioned in Sect.~\ref{sec:line_int} can be found in Table~\ref{tab:line_fluxes_h2o}. We note that these are only normalised to the common distance, but not scaled by $L_\text{acc}$. The average of the \ce{H2O} line fluxes of the different energies and their ratio can be found in Fig.~\ref{fig:water_radius_corr}. When discussing these quantities, we shall refer to them as the  \ce{H2O}$_\text{cold}$ or \ce{H2O}$_\text{warm}$ line fluxes. In the case of SY~Cha and PDS~70, some of the higher energy lines are not detected, and therefore an upper limit is shown. A more detailed view of the gas emission in PDS~70 will be presented in Dohrman et al. (in prep.). Overall, DR~Tau shows the strongest \ce{H2O} emission in both the warm and cold lines of the full discs (SY~Cha shows stronger cold \ce{H2O} emission), though not the strongest \ce{H2O}$_\text{cold}$/\ce{H2O}$_\text{warm}$ ratio. BP~Tau has the strongest \ce{H2O}$_\text{cold}$/\ce{H2O}$_\text{warm}$ ratio out of the full discs where all the selected \ce{H2O} lines are detectable. BP~Tau is also the smallest disc in the sample, and may have a small inner cavity (see Fig.~\ref{fig:radial_profiles}). The weakest \ce{H2O} fluxes correspond to V1094~Sco. We note that the \ce{H2O}$_\text{cold}$/\ce{H2O}$_\text{warm}$ of CI~Tau and IQ~Tau compared to the smaller discs in this sample appears to agree with \citet{ref:23BaPoCa}, where they are also included. 

In Fig.~\ref{fig:water_radius_corr}, the PCCs and p values are given for a correlation with $R_{\text{dust}}$ (excluding PDS~70 and SY~Cha). When looking at the general trends in Fig.~\ref{fig:water_radius_corr}, similarly to \citet{ref:20BaPaBo} and \citet{ref:23BaPoCa}, we observe that the overall warm and cold line fluxes decrease with outer dust disc radius, along with the ratio between the fluxes covering the two energies. However, none of these are perfectly straight lines, and only a moderate correlation is observed in this sample due to Sz~98 and V1094~Sco. If one were to also disregard the two most luminous objects in the sample, the ratio of the line fluxes would almost be a straight line. Similarly, the p values indicate some statistical significance of these results, but the sample is too small to be certain.

In Fig.~\ref{fig:water_radius_corr}, there are four discs with a higher \ce{H2O}$_\text{cold}$/\ce{H2O}$_\text{warm}$ ($E_{\rm{up}} < 4000$~K / $6000 \leq E_{\rm{up}} < 8000$~K) than DR~Tau: BP~Tau, Sz~98, SY~Cha, and PDS~70, although we note that the latter two are lower limits, since the warm lines shortward of $\sim$16.5~$\mu$m are not detected. Looking at the comparison with the spectral template in Fig.~\ref{fig:water_excess} in the bottom four panels, it indeed becomes clear that the colder lines at longer wavelengths (right column) are stronger than that of DR~Tau. However, two of these discs also show a clear difference in the shape of the 23.8 to 24~$\mu$m \ce{H2O} quadruplet, where the first and third lines in the set trace the lowest energies \citep{ref:24TeDiGa}. This is the case for Sz~98 and SY~Cha, where the second and fourth lines have the same height as DR~Tau, but the first and third are stronger.
This excess in the coldest component of \ce{H2O} emission indicates that these two discs may be more enhanced by cold \ce{H2O} at the snowline due to drifting icy pebbles than DR~Tau. For the other two, BP~Tau and PDS~70, the shape of the quadruplet is largely similar, meaning that the warm and cold components of the spectrum have similar emitting properties compared to DR~Tau.

\begin{figure}[t]
    \centering
    \includegraphics[width=0.6\columnwidth]{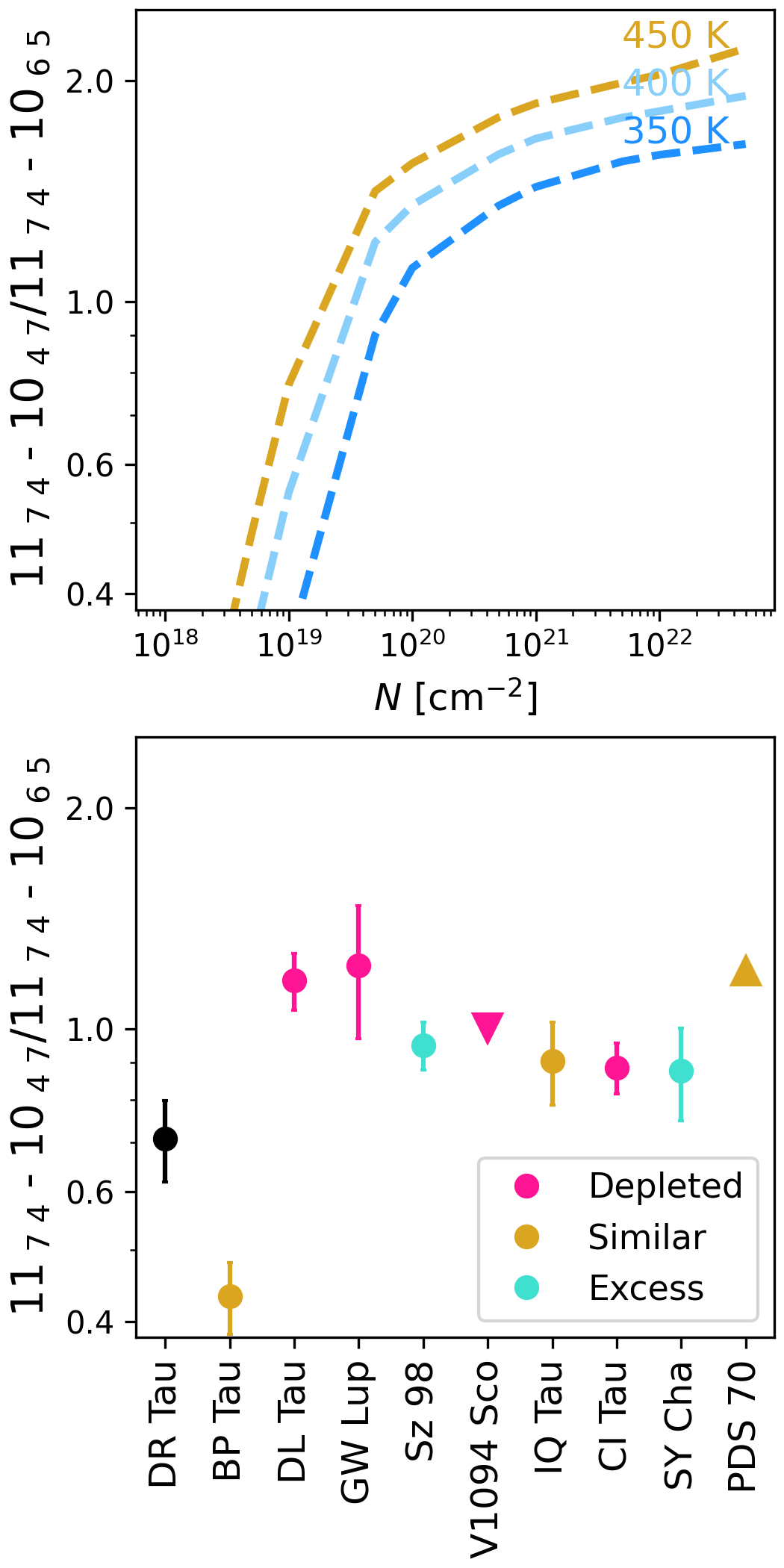}
    \caption{Line flux ratio of the column density tracer from slab models (top), and the same line pair ratio in the spectra (bottom). The colour-coding corresponds to the observed depletion and excesses in colder \ce{H2O} compared to DR~Tau (black point), where similar emission properties to DR~Tau are indicated with gold markers, an excess cyan markers, and a depletion pink markers. In the top plot, the temperature effects on the line flux ratio are demonstrated. In V1094~Sco and PDS~70 we do not detect one of the lines of the cold pair. Therefore, the lower/upper limits are indicated by the triangles in the bottom panel.}
    \label{fig:cold_col_corr}
\end{figure}

The remaining five discs (top five panels in Fig.~\ref{fig:water_excess}) show some variety in the quadruplet. Aside from BP~Tau and PDS~70, the emission of IQ~Tau is also similarly shaped, with the first and third lines being stronger. However, the overall emission here is weaker than in DR~Tau, as opposed to BP~Tau and PDS~70. The final four discs, V1094~Sco, DL~Tau, GW~Lup, and CI~Tau are clearly depleted in the cold \ce{H2O} (or enhanced in warm \ce{H2O}) compared to DR~Tau. Based on Fig.~\ref{fig:water_radius_corr}, V1094~Sco does show a somewhat larger \ce{H2O}$_\text{cold}$ excess compared to the other depleted discs, and also appears the least depleted of the four in Fig.~\ref{fig:water_excess}, but overall its \ce{H2O} emission is the lowest.

In Fig.~\ref{fig:cold_col_corr} we present the flux ratio of the low-energy line pair (11$_{\text{ }7\text{ }4}$--10$_{\text{ }4\text{ }7}$                                 and 11$_{\text{ }7\text{ }4}$--10$_{\text{ }6\text{ }5}$). Here, the discs are colour-coded depending on whether the relative emission within the quadruplet is observed to show similar relative fluxes to DR~Tau (gold), the quadruplet indicates an excess compared to the warm \ce{H2O} and DR~Tau (cyan), or a depletion (pink), as was identified previously in this section. In addition, the top panel demonstrates how the flux ratio of this line pair changes with column density and temperature, where it is important to keep in mind that they may be populated differently with changes in temperature. It shows that aside from a higher column density, a higher temperature also increases the ratio between the lines.

The discs that show a clear excess in the colder emission both have a larger line flux ratio than DR~Tau. Keeping in mind that the 11$_{\text{ }7\text{ }4}$--10$_{\text{ }4\text{ }7}$ transition is likely not optically thick and traces a warmer column of approximately 400~K in DR~Tau, the flux ratio indicates that all discs, except BP~Tau, have a higher column density of \ce{H2O}, even those that appear depleted in the very cold \ce{H2O} component. Only BP~Tau has a lower ratio, indicating a lower column density.

\begin{figure*}[ht]
	\centering
    \includegraphics[width=\textwidth]{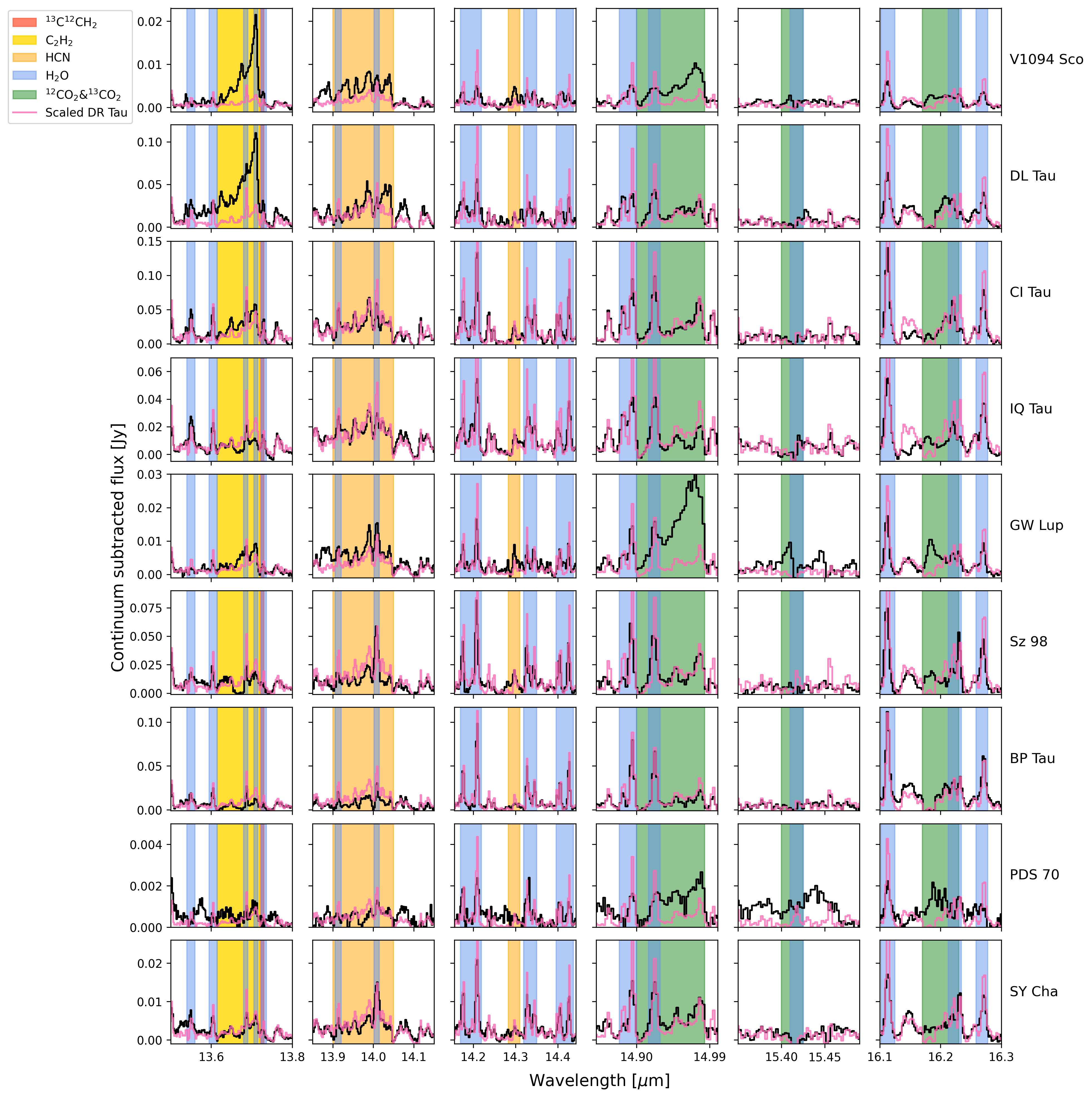}
    \caption{Continuum-subtracted spectra, zoomed in on \ce{C2H2} and $^{13}$C$^{12}$CH$_2$ (left-most column), the \ce{HCN} $Q$-branch and hot branch (second column and third column), the \ce{CO2} $Q$-branch, \ce{^{13}CO2}, and \ce{CO2} hot branch (fourth, fifth, and sixth column). The respective features are highlighted in yellow, orange, or green, and prominent \ce{H2O} features are highlighted in blue. The spectra are once more ordered from largest to smallest \ce{H2O}/$L_\text{acc}^{0.6}$ from bottom to top. Note that in the fifth column, the highlighted \ce{H2O} feature has been subtracted to uncover a potential \ce{^{13}CO2} feature. The rescaled spectrum of DR~Tau is included in pink for comparison.}
    \label{fig:carbon_iso}
\end{figure*}

\begin{figure}[t]
    \centering
    \includegraphics[width=\columnwidth]{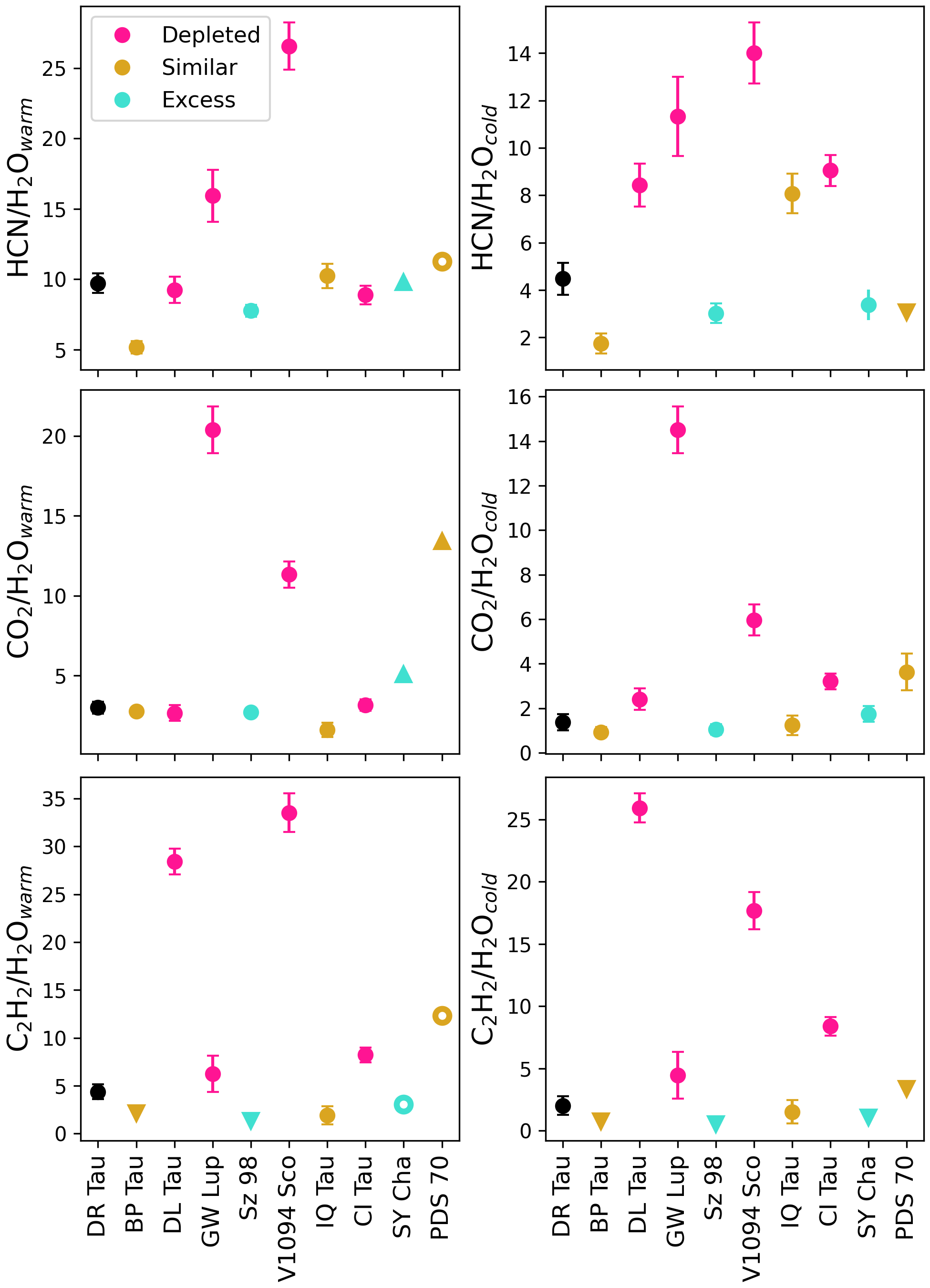}
    \caption{Ratio of line fluxes of carbon-bearing molecules and warm (left) and cold (right) H$_2$O. The colour coding follows that of Fig.~\ref{fig:cold_col_corr}: the discs indicated by a gold marker show similar cold \ce{H2O} emission as DR~Tau, cyan an excess, and pink a depletion. An open marker indicates a non-detection in both the carbon and \ce{H2O} lines examined, whereas the triangles indicate lower or upper limits due to non-detections.}
    \label{fig:carbon_R}
\end{figure}

\subsubsection{C$_2$H$_2$}
In the following sections, we compare the emission of the carbon species between rescaled DR~Tau and the other discs. This comparison can be found in Fig.~\ref{fig:carbon_iso}. First, we look at the left-most column, showing the \ce{C2H2} and $^{13}$C$^{12}$CH$_2$ emission, highlighted in yellow.

DR~Tau itself shows clear \ce{C2H2} emission, but its isotopologue is dominated by an \ce{H2O} line. It has neither the strongest, nor the weakest \ce{C2H2} emission in the sample, compared to the warm \ce{H2O} emission. IQ~Tau, Sz~98, and BP~Tau are weaker in \ce{C2H2}. SY~Cha and potentially PDS~70 (although this source is quite noisy and \ce{C2H2} is not considered detected) are similar in their \ce{C2H2}. CI~Tau and GW~Lup on the other hand, are slightly elevated compared to DR~Tau\, though this difference is very small. Finally, both V1094~Sco and DL~Tau show significant enhancement in \ce{C2H2}. The $^{13}$C$^{12}$CH$_2$ feature, highlighted in yellow, overlaps with \ce{H2O}, but the emission from \ce{H2O} is weaker in these discs. Therefore, the emission in this region is more likely to originate from $^{13}$C$^{12}$CH$_2$, indicating the \ce{C2H2} emission is not just elevated, but also high in column density.

\subsubsection{HCN}
The second and third columns show the \ce{HCN} $Q$-branch and the hot branch. Again, DR~Tau is neither exceptionally strong nor exceptionally weak compared to the other discs. Discs that are weaker in \ce{HCN} are Sz~98, BP~Tau, and to a lesser degree PDS~70 (once more limited by noise) and SY~Cha. IQ~Tau, GW Lup, and CI~Tau are the same as DR~Tau, whereas DL~Tau and V1094~Sco are elevated in \ce{HCN}. However, the \ce{HCN} feature is blended with \ce{C2H2}, so for DL~Tau and V1094~Sco some of the peaks are affected by the strong \ce{C2H2} present in the spectrum. However, for the integrated line fluxes, this is mitigated by subtracting \ce{C2H2} slab model fits.

Overall, a strong detection of \ce{C2H2} coincides with detectable \ce{HCN}, but \ce{HCN} is more widely detectable than \ce{C2H2}. For most discs, the \ce{HCN} feature is stronger than \ce{C2H2}, except for the very carbon-rich DL~Tau and V1094~Sco, where \ce{C2H2} is brighter.


\subsubsection{CO$_2$}
Finally, the three columns on the right of the figure demonstrate the emission of the \ce{CO2} $Q$-branch, $^{13}$CO$_2$, and the \ce{CO2} hot branch, respectively. The detection of $^{13}$CO$_2$, and a \ce{CO2} hot branch with a relatively bright peak compared to the main $Q$-branch shortward of 16.2~$\mu$m are indicators for elevated column densities (clear examples of both can be seen in GW~Lup, as has also been discussed by \citealt{ref:23GrDiTa}).

In terms of the $Q$-branch, several sources appear similar to DR~Tau: SY~Cha, BP~Tau, Sz~98, CI~Tau, and DL~Tau. Sz~98 may have an enhanced column density, since its spectrum shows stronger emission from the hot branch compared to the main $Q$-branch. Similarly, IQ~Tau is weaker in the $Q$-branch, but stronger in the hot branch, which indicates that its \ce{CO2} emission originates from a smaller emitting area, but with a higher column density. Finally, the other sources are enhanced in \ce{CO2}. These include PDS~70 (although its spectrum is noisy due to the faintness of the source), GW~Lup, and V1094~Sco. For the latter, the $^{13}$CO$_2$ seems to also be detectable, indicating a high column density, similarly to GW~Lup.

\subsubsection{Carbon versus H$_2$O}
The comparisons above give an idea of the emitting strength of the carbon-bearing species compared to the warm \ce{H2O}, since that is the measure on which the DR~Tau rescaling is based. However, a comparison to the cooler \ce{H2O} reservoir is also of interest, in particular whether a trend can be seen between an excess of cooler \ce{H2O}, a potential indication for \ce{H2O}-ice transportation, and the strength of the carbon emission. The line fluxes of the carbon-bearing species measured in the manner discussed in Sect.~\ref{sec:line_int} can be found in Table~\ref{tab:line_fluxes_carbon}. Similarly to the \ce{H2O} line fluxes, these are only normalised to the common distance, but not scaled by $L_\text{acc}$. The comparison to the \ce{H2O} line fluxes is shown in Fig.~\ref{fig:carbon_R}, where the same colour-coding is used as in Fig.~\ref{fig:cold_col_corr}. The left column shows the flux ratios with \ce{H2O}$_\text{warm}$, and the right column the ratios with \ce{H2O}$_\text{cold}$. The windows over which the fluxes are calculated are mentioned in Sect.~\ref{sec:line_int}.

First, no dichotomy appears in the ratios with \ce{H2O}$_\text{warm}$ between the sources that are depleted or enhanced in \ce{H2O}$_\text{cold}$. Generally, only two of the depleted sources are stronger in carbon emission compared to the warmer \ce{H2O}: DL~Tau and V1094~Sco.

On the other hand, a split occurs between the samples for the ratio with colder \ce{H2O}. The depleted sources are stronger in carbon, especially \ce{HCN} and \ce{C2H2}, while the other discs are more centred near the bottom of the plots. The two sources that appear even more enhanced in \ce{H2O}$_\text{cold}$ also have very weak, if at all detectable, \ce{C2H2}. The only sources that are not always split well from the depleted sources are PDS~70, which is not a full disc, and IQ~Tau, which appears relatively strong in \ce{HCN}. The latter could also be due to the weak colder \ce{H2O} in IQ~Tau. However, in general the discs depleted in cold \ce{H2O} show stronger emission from carbon-bearing species.

\section{Discussion}
\label{sec:discussion}

\begin{table*}[t]
\caption{Summary of the emission per disc, mostly compared to DR~Tau.}
\label{tab:emission_summary}
\centering
\resizebox{1\linewidth}{!}{%
\begin{tabular}{lllllllll||llcc} \hline \hline
ID        & \ce{H2O}$_\text{hot}$ & \ce{H2O}$_\text{cold}$ & \multicolumn{1}{c}{\begin{tabular}[c]{@{}c@{}} \ce{CO2}\\ ($Q$-branch)  \end{tabular}} & \multicolumn{1}{c}{\begin{tabular}[c]{@{}c@{}} \ce{CO2} \\ ($N$) \end{tabular}} & \multicolumn{1}{c}{\begin{tabular}[c]{@{}c@{}} \ce{HCN} \\ ($Q$-branch) \end{tabular}} &  \multicolumn{1}{c}{\begin{tabular}[c]{@{}c@{}} \ce{HCN} \\  (hot) \end{tabular}} & \multicolumn{1}{c}{\begin{tabular}[c]{@{}c@{}} \ce{C2H2} \\ ($Q$-branch) \end{tabular}} & $^{13}$C$^{12}$CH2 & C & O & \multicolumn{1}{c}{\begin{tabular}[c]{@{}c@{}}Inside\\ snowlines? \end{tabular}} & \multicolumn{1}{c}{\begin{tabular}[c]{@{}c@{}}Scenario\\ younger (y), older (o) \end{tabular}}\\ \hline
V1094 Sco & Low     &  \cellcolor{red} &  \cellcolor{green}  & \cellcolor{green} & \cellcolor{green} & \cellcolor{green} & \cellcolor{green} & \cellcolor{green} &  \cellcolor{green} & \cellcolor{red} & Yes & 4 (o) \\
DL Tau    & Low     &  \cellcolor{red} &  \cellcolor{yellow} & \cellcolor{yellow} & \cellcolor{green} & \cellcolor{yellow} & \cellcolor{green} & \cellcolor{green} & \cellcolor{green}  & \cellcolor{red} & Yes & 1, 3, 5 (o) \\
GW Lup    & High  &  \cellcolor{red} & \cellcolor{green}  & \cellcolor{green} & \cellcolor{yellow} & \cellcolor{green} & \cellcolor{green} & \cellcolor{yellow} &  \cellcolor{green} &  \cellcolor{red} & No & 3, 4 (o) \\
CI Tau    & High    &  \cellcolor{red} &  \cellcolor{yellow} & \cellcolor{yellow} & \cellcolor{yellow} & \cellcolor{yellow} & \cellcolor{green} & \cellcolor{yellow} & \cellcolor{yellow}  &  \cellcolor{red} & Yes & 4 (o) \\
IQ Tau    & High    &  \cellcolor{yellow} & \cellcolor{red} & \cellcolor{yellow}  & \cellcolor{yellow} & \cellcolor{yellow} & \cellcolor{red} & \cellcolor{yellow} & \cellcolor{yellow}  &  \cellcolor{yellow} & No & 1, 2 (o) \\
Sz 98     & Medium  &  \cellcolor{green}  & \cellcolor{yellow} & \cellcolor{green} & \cellcolor{yellow} & \cellcolor{yellow} & \cellcolor{red} & \cellcolor{yellow} & \cellcolor{yellow}  & \cellcolor{green} & Yes & 2 \\
BP Tau    & High    &  \cellcolor{yellow} & \cellcolor{red} & \cellcolor{yellow} & \cellcolor{red}  & \cellcolor{yellow} & \cellcolor{red} & \cellcolor{yellow} &  \cellcolor{red} & \cellcolor{yellow} & - & 1 \\
PDS 70    & Low     &  \cellcolor{yellow} & \cellcolor{green} & \cellcolor{green} & \cellcolor{red} & \cellcolor{yellow} & \cellcolor{red} & \cellcolor{yellow} &  \cellcolor{red} &  \cellcolor{yellow} & - & 1-5 \\
SY Cha    & High    &  \cellcolor{green}  & \cellcolor{yellow} & \cellcolor{yellow} & \cellcolor{yellow} & \cellcolor{yellow} & \cellcolor{yellow} & \cellcolor{yellow} & \cellcolor{yellow}  & \cellcolor{green} & - & 1-5 \\ \hline
\end{tabular}%
}
{\raggedright \textbf{Notes.} The \ce{H2O}$_\text{hot}$ is based on the sample as a whole, not DR~Tau alone. Green indicates larger fluxes than rescaled DR~Tau, yellow approximately the same, and red less than DR~Tau. The $N$ for \ce{CO2} indicates the relative \ce{CO2} column density based on the $^{13}$CO$_{2}$ and the \ce{CO2} hot branch. The right-most column indicates whether the substructures are present within the \ce{CO} and \ce{CH4} snowlines. \par}

\end{table*}

\begin{figure*}[ht]
	\centering
    \includegraphics[width=\textwidth]{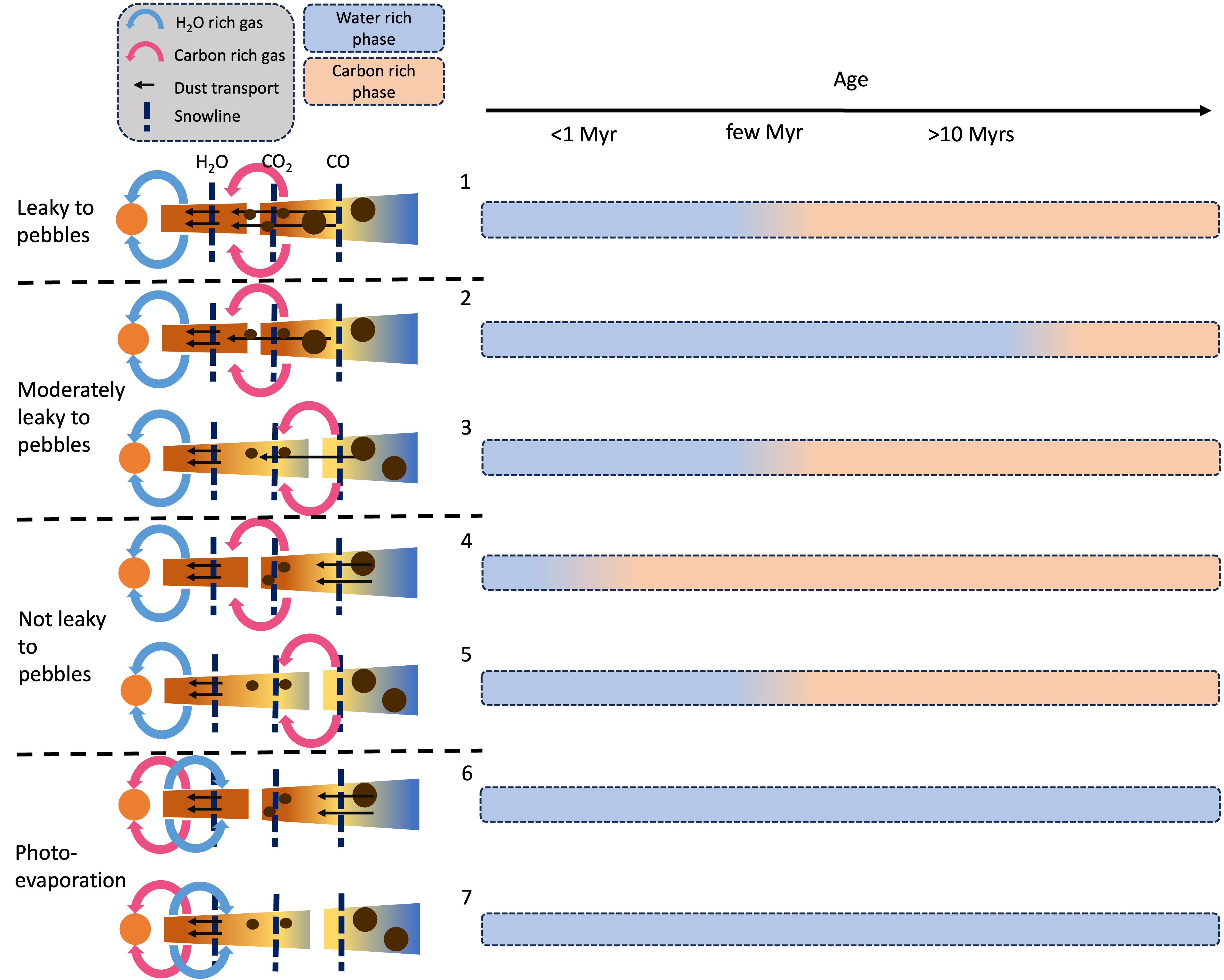}
    \caption{Emission properties of the inner disc based on the modelling works by \citet{ref:21KaPiKr}, \citet{ref:23KaPiKr}, \citet{ref:24MaSaBi}, \citet{ref:24LiBiHe}, and \citet{ref:24SeVlDi}.}
    \label{fig:scenarios}
\end{figure*}

\subsection{Inner disc emission and outer dust disc structure}
\label{sec:disks}
In Sect.~\ref{sec:results}, we discussed the inner disc emission of the sample. A summary of the results can be found in Table~\ref{tab:emission_summary}. Despite the presence of substructures in all discs, some discs are \ce{H2O} dominated and show signs of active dust migration, whereas others are much more carbon-rich. Based on modelling works, we identify different scenarios and try to put the observed emission into context, and illustrate these in Fig.~\ref{fig:scenarios}. Before we discuss the discs in this sample in the context of the diagram, we first take the reader through the different scenarios based on the different models considered. We note that in all of these, a two population dust distribution is assumed, where some of the small dust is allowed to leak through the gap, independent of what happens to the pebbles. In this context, assumptions are made related to how much ice is carried by these small grains and how much can leak through the gaps, both of which are likely underestimated due to the simplified representation of the dust population. The scenarios discussed below and presented in Fig.~\ref{fig:scenarios} are directly based on these models, but this caveat must be kept in mind.

If the substructures are very leaky to dust pebbles, and therefore shallow, we can expect the disc evolution to progress similarly to a full, gap-free, viscous disc \citep[e.g.][]{ref:19BoIl,ref:24MaSaBi}. This corresponds to scenario 1, where ices on dust pebbles are drifting in first, where ice species subsequently sublimate into the gas at their respective snowlines and no longer migrate inwards at the same rate as the pebbles. Therefore, the inner disc first becomes enriched in \ce{H2O} (and we expect to be able to observe colder \ce{H2O} out to the snowline), after which gases from the outer disc find their way inwards and the disc becomes enriched in \ce{CO2}, and subsequently other carbon-bearing species. The majority of the \ce{H2O} emission may appear hotter as it is accreted onto the star.

If a gap is only somewhat leaky to dust pebbles, it is a 'moderate' gap, and the inflow of dust is more regulated. This corresponds to scenarios 2 (close-in gap) and 3 (far out gap) in the figure, where some flux of pebbles is still able to cross the gaps. In the models of \citet{ref:24MaSaBi}, this results in significantly prolonged \ce{H2O}-enrichment phases (and colder \ce{H2O}), since the rate at which the dust moves into the inner disc is somewhat reduced. The outer disc would not run out of these grains as quickly. Based on \citet{ref:21KaPiKr,ref:23KaPiKr}, and \citet{ref:24SeVlDi}, we observe that a deep gap closer to the star is more effective at blocking dust pebbles compared to scenario 3, as there simply is more dust starting outside of the gap. Therefore, the difference between scenarios 2 (close-by gap) and 3 (gap farther away) is that in scenario 2 the \ce{H2O} enrichment period may last longer compared to scenario 3, longer than 10~Myrs in the models of \citet{ref:24MaSaBi}.

If the gap is capable of blocking most, if not all, of the dust pebbles, it is a deep gap. In the models discussed, this can be caused by either planet formation-like mechanisms, or photoevaporation. The former is shown in scenarios 4 (close-in gap) and 5 (far out gap). In these cases, gas is allowed to diffuse across the gaps, but pebble drift is completely prevented. Realistically, in these cases the strong pressure bump can fragment the pebbles into smaller dust grains, in which case the dust can continue to migrate with the gas. If the gap is then located closer to the star, the dust reservoir inside the radius of the gap is quickly depleted, and the \ce{H2O} enrichment phase is very short. The visible colder \ce{H2O} emission quickly disappears, perhaps after less than a million years \citep{ref:24MaSaBi}. On the other hand, a gap farther from the star is capable of blocking less dust, and the enrichment phase would last longer.

If photoevaporation processes are the cause of the gap, small dust grains and gas are also prevented from crossing the gap. This is illustrated in scenarios 6 (inside \ce{CO2} snowline) and 7 (outside \ce{CO2} snowline). Photoevaporative winds blow the gas out of the gap \citep{ref:24LiBiHe}. Contrary to all aforementioned scenarios, the cold \ce{H2O} reservoir is continuously replenished. The gas in the inner disc diffuses outward towards the photoevaporative zone. If this zone lies outside the \ce{H2O} snowline, but inside the snowlines of \ce{CO}, \ce{CO2}, and \ce{CH4}, the \ce{H2O} gas is able to condense back onto the dust and is carried inwards, while the other gases are lost to the photoevaporative zone \citep{ref:24LiBiHe}. However, this phase quickly results in a large cavity, depending on the viscosity of the disc, and occurs towards the end of the gas disc's lifetime. Therefore, while an important scenario to keep in mind, it is unlikely that any of the discs presented here are exactly in the gap-opening phase.


An extended discussion of the individual discs in the context of Fig.~\ref{fig:scenarios} is included in App.~\ref{app:extended_discussion}. The most important points are summarised here. Of the discs that show no cold \ce{H2O} excess (see Table~\ref{tab:emission_summary}), only two discs (GW~Lup and IQ~Tau) have no gap inside the snowlines. For these discs we find evolved versions of scenarios 3 and 5 more likely. In contrast, if the substructures closer to the star are indeed deep enough to effectively block dust pebbles, V1094~Sco, DL~Tau, and CI~Tau may be examples of an evolved scenario 4. Of the sources with some cold \ce{H2O} emission, DR~Tau and Sz~98 show substructures within the snowlines. These discs may be examples of scenario 2, where the close gaps are not blocking all incoming dust pebbles. Since Sz~98 has a larger excess in cold \ce{H2O} than DR~Tau, more \ce{H2O}-ice may be crossing the snowline. Furthermore, Sz~98 is much weaker in \ce{C2H2} and \ce{HCN} compared to the other discs, which may correspond to scenario 6, though, as was mentioned above, it is unlikely to observe a system exactly in the gap-opening phase. IQ~Tau falls somewhere between the cold \ce{H2O} enhanced and depleted discs, and may be more evolved versions of scenarios 1 and 2.

The substructure identified in BP~Tau is a cavity of a few astronomical units. Such a cavity may influence the inner disc emission, either due to gas depletion \citep[e.g.][]{ref:17BaPoSa} or heating of the inner cavity wall \citep{ref:24VlDiTa}. In this case, the snowlines may be shifted outwards, resulting in enhanced \ce{H2O} flux. Due to the cavity, it is difficult to place BP~Tau in the context of Fig.~\ref{fig:scenarios}. However, it may act most closely to scenario 1, since it lacks in detectable strong substructures aside from the cavity.

Discs with a very deep, extended gap (PDS~70 and SY~Cha), are stronger in \ce{H2O} emission than the carbon species. \ce{H2O} is still present in PDS~70 and SY~Cha, which indicates that some gas and dust is still reaching the inner disc across their large deep gaps. This is especially interesting in the case of SY~Cha, which shows a stronger excess in cold \ce{H2O} than DR~Tau, indicating its emission is more drift-dominated, which is in line with the detection of gas within its gap \citep{ref:23OrMoMu}. This dust may not be larger pebbles, as is typically discussed in the models and when considering dust drift, but smaller dust grains moving inwards along with the gas. These smaller dust grains may still carry enough ices to replenish the inner disc in \ce{H2O}, and other species, as is also discussed in \citet{ref:23PeChHe}, \citet{ref:24PiBeWa}, and \citet{ref:24JaWaKa}. In this case all scenarios from 1 to 5 may be valid, since small dust rather than pebbles bring the ices to the inner disc. Whether this is more universally happening in this type of disc, or whether SY~Cha and PDS~70 are exceptions remains to be seen. An overview of the gas emission in transition discs will be available in Perotti et al. (in prep.).

Evidence for a variety of scenarios is found, but more data is needed to identify whether gaps truly have a significant impact on the inner disc composition, and which scenarios are most prevalent. Regardless, \ce{H2O}-rich scenarios exist even in the presence of seemingly deep gaps, indicating that these substructures do not seem capable of fully blocking replenishment by ices or gases. This relates back to the aforementioned caveat, and it is important to consider the contribution of small dust to the ice being brought into the inner disc in the future. Furthermore, there may be differences in how much replenishment happens. Constraining the exact abundance of the gas species in the inner disc (by means of e.g. H$_2^{18}$O detections, or optically thin lines with lower $A_{ul}$) is therefore paramount.

\subsection{Volatile trapping in ices}
The calculation of the snowlines often assumes no volatile trapping within the ices and of ices within refractories \citep{ref:24PoJaMu}, which results in the classical interpretation of the C/O ratio changes in different parts of the disc \citep[e.g.][]{ref:11ObMuBe}. However, ices are likely not neatly organised and separated on the dust particles, and may be trapped within ices of other species that have their snowlines closer to the star, as is seen in protostellar envelopes \citep[e.g.][]{ref:15BoGeWh} and the edge-on disc HH~48~NE \citep{ref:23StMcBe}. In the context of this work, trapping of volatiles within \ce{H2O} ice is of particular interest, since this would make the locations of other snowlines relative to substructures of smaller influence, since a fraction of the volatiles can be trapped within ices. Therefore, when discussing volatile trapping, we mean trapping of volatiles within \ce{H2O} ice. 

This is examined by \citet{ref:24LiKiGa} and Bergner et al. (subm.) for discs, and causes the gaseous C/O ratio to be much less dependent on the radial location within the disc, up to the \ce{H2O} snowline where most volatiles are released from the \ce{H2O} ice. The volatile trapping was studied in a lab context by \citet{ref:04CoAnCh}, showing that a large fraction of the volatiles may be released at several temperature regimes. Certain species, such as the hydrocarbons including \ce{CH4}, but also \ce{CO2}, are expected to be mixed in with the \ce{H2O} ice, moving their true desorption temperature to that of \ce{H2O} ice, as they cannot be released unless the surrounding \ce{H2O} is. Additionally, \ce{H2O} ice may be trapped in refractories and can only be released at the 400~K sublimation temperature of, for example, silicates \citep[e.g.][]{ref:24PoJaMu}. On the other hand, \ce{CO} ice is more apolar \citep{ref:06Po}, and likely a large fraction of the \ce{CO} ice is free to be released into the gas at its own sublimation temperature \citep{ref:04CoAnCh}.

While \ce{CO} may not be efficiently trapped, it can be chemically processed into other ices on the timescale of a few million years in discs around T~Tauri stars \citep[e.g.][]{ref:18BoWaDi,ref:20KrBoZh}. In the cold environment of the outer disc, \ce{CO} ice is processed into \ce{CO2} and \ce{CH3OH} ice. Especially the former two ice species reach their sublimation temperatures much closer to the star, similarly to \ce{H2O}, resulting in a similar effect as volatile trapping of \ce{CO} itself would have.

This indicates that, similarly to the discussion in \citet{ref:24LiKiGa}, the composition of the gas and solids can be relatively constant throughout the outer disc, until the \ce{H2O} and \ce{CO2} snowlines are reached. The location of the gaps relative to the snowlines, as was previously calculated, may therefore be of little consequence, unless they are inside the \ce{H2O} and \ce{CO2} snowlines. A large part of the major carbon carriers is either trapped in \ce{H2O} ice and brought in at the same time as \ce{H2O}, or reprocessed into \ce{CO2} and \ce{CH3OH}. Rather than reaching the inner disc later due to slower transport of gases, as is typically assumed for viscous, full disc evolution, they remain attached to the dust for longer and reach the inner disc simultaneously. Some progression from an oxygen-rich towards a more carbon-rich inner disc may still occur, since the outer disc gases that are there do tend to be more carbon-rich, but the notion of more carbon-rich species being left behind at their respective snowlines becomes less applicable.

The discs with the \ce{CO} and \ce{CH4} snowlines outside some of their detectable substructures, like DR~Tau, Sz~98, CI~Tau, DL~Tau, and V1094~Sco, may therefore experience not only the regulation in the influx of \ce{H2O}, but in fact of many of the carbon species as well, just like the discs with the snowlines within their detectable substructures. The location of gaps with respect to how much material it can block in general (e.g. how much dust is present at larger radial distances) would matter more. In this case, a detectable colder \ce{H2O} reservoir as a drift tracer would coincide with elevated emission in the other species as well. Therefore, the presence of a \ce{H2O}$_\text{cold}$ excess, in addition to considerable emission from the other molecular species discussed here, could be an indication of relatively ‘free' viscous disc evolution. The lack of this \ce{H2O}$_\text{cold}$ excess, but clear emission from other species would indicate that most material being brought in is from the gas, and subsequently reprocessed through chemical reactions depending on the inner disc conditions, or very little radial transport occurs at all. The presence of an \ce{H2O}$_\text{cold}$ excess, but lack of emission from other species, would instead have to be a consequence of other processes dominating the inner disc emission, such as photoevaporation, the presence of a cavity, the entire disc having significantly different volatile ratios, or less trapping of ices.

\subsection{Ages and formation timescales}
For later stages of evolution to be reached, for example a more carbon-rich inner disc, the disc must be older, or the evolution timescale (as is dictated by the disc viscosity) relatively short. Constraining these properties would allow us to get a better understanding of what scenarios in Fig.~\ref{fig:scenarios} are more applicable. The inner disc composition can appear similar no matter how effective a gap is at blocking dust, if the evolution timescale is sufficiently slow, or the disc sufficiently young. As is modelled in \citet{ref:24MaSaBi} and \citet{ref:24SeVlDi}, the more viscous the disc, the faster its transport, and the faster different enhancement phases pass. Furthermore, more viscous discs contain more small dust grains, and substructures will generally be less effective at fully blocking dust, resulting in prolonged enrichment scenarios. On top of dust being less effectively blocked, the gas would also be transported inwards faster \citep{ref:24MaSaBi}. If the silicate dust features in Fig.~\ref{fig:all_specs} are any indication of the abundance of small dust grains \citep{ref:06KeAeDu}, the more peaked dust features exhibited by DR~Tau, BP~Tau, Sz~98, PDS~70, and SY~Cha would hint at more small dust reaching the inner disc. This could be an indication of a higher disc viscosity and subsequently leakier gaps. The discs would then have to be relatively young to be able to retain the \ce{H2O} excess they exhibit currently. However, dust grain sizes and the dust opacity are not stagnant and only dependent on the dust that is drifting inwards. These features may not allow us to distinguish between dust processing as is a natural occurrence for the discs, and the kind of dust that is reaching the inner disc.

Furthermore, as is also noted by \citet{ref:24MaSaBi}, for gaps to have a significant influence on the inner disc composition, they must form early in the disc lifetime. They propose that very deep gaps due to growing planets may only be possible to form early enough in very massive discs. Some of the more massive discs in this sample include V1094~Sco and DL~Tau, both of which are poor in \ce{H2O}. If their gaps are deep and did form early, this could have resulted in a quick passage of the \ce{H2O} enriched phase resulting in the carbon-dominated discs we see now. However, planet formation is not the only mechanism that can carve gaps, and as has already discussed, the nature of the gap differs depending on what caused its formation \citep{ref:23BaIsZh,ref:24LiBiHe}, although photoevaporation may only occur at very late stages of disc evolution.

\subsection{Dust opacity}
The hypothesis that disc substructures in the outer disc influence the inner disc composition, relies on transport of dust. Therefore, along with a delivery of oxygen-rich ices like \ce{H2O}, dust is also moving into the inner disc. The consequences of this are discussed in \citet{ref:24SeVlDi}, who show that this can also result in an increase in dust opacity, assuming the dust is coupled to the gas and a `traffic jam' of grains occurs inside the \ce{H2O} snowline. This increase in dust opacity may block the emission from gases closer to the midplane, resulting in the same column density of \ce{H2O} gas emitting above the dust compared to before the grains drifted inwards, based on retrievals using synthetic spectra. They posit that the \ce{H2O} emission properties and column density as probed by the mid-IR lines is a poor tracer of how much gas is actually present in the inner disc, due to this dust opacity increase. Even if the gas and dust are not coupled, the discs in the sample may have differences in dust opacity in the inner disc.


There have been some indications from models for certain species preferentially forming in different vertical layers of the disc \citep[e.g.][]{ref:18WoMiTh}. Depending on where the dominant reservoirs of the different species are located, the differences in dust opacity of the discs in this sample could be influencing the emission that is visible in the MRS spectrum. More gas closer to the midplane could be obscured for one case, whereas another might be more transparent. The inner disc dust composition and abundance could therefore influence the species visible in the spectrum. Overall, the discs with flat-topped 10~$\mu$m silicate features (CI~Tau, GW~Lup, DL~Tau, and V1094~Sco) show stronger emission from carbon-bearing species such as \ce{C2H2} and \ce{HCN}, which would be in line with these molecules being located closer to the midplane than \ce{H2O}.

\subsection{Gas-to-dust ratio}
Works that do not include migration of ices in their models are also able to reproduce the relative line fluxes of \ce{H2O} and other species, such as \ce{HCN}. The works of \citet{ref:15AnKaRi} and \citet{ref:18WoMiTh}, for example, show that the easiest way to achieve this is by increasing the gas-to-dust ratio or gas mass, where the former works somewhat like a reduced dust opacity, and the latter simply dictates whether there is enough gas present to produce detectable emission lines. Additionally, a more flared disc can be heated more by the central star, resulting in warmer and somewhat stronger emission from \ce{H2O} in the MIRI/MRS wavelengths \citep{ref:15AnKaRi}. The migration and processing of dust in general may be sufficient to produce enhancements of \ce{H2O} fluxes and most other species, independent of the ices brought in by the dust \citep{ref:23AnKaWa}. In this case the \ce{H2O} enhancement would still be age-dependent, but not dependent on the outer disc ice reservoir available for replenishment. This once more stresses the importance of accurate age estimates. 

\citet{ref:19GrKaWa} find \ce{C2H2} to not be affected by dust processing as strongly (demonstrated by their Fig.~9), even though we find a general increase in \ce{C2H2} flux for the discs with weaker \ce{H2O}. \citet{ref:19GrKaWa} also note this discrepancy with previous results from observations, and that their \ce{C2H2} emission behaves very differently from the emission of other species. They mention that these differences may be caused by missing formation pathways in the chemical network, or that their assumed dust models do not adequately describe the dust in \ce{C2H2}-rich discs. \citet{ref:23AnKaWa} further argue that additional enrichment due to ices should not significantly affect the emission lines of \ce{H2O} originating from the inner disc, which are in general optically thick already. However, the models in \citet[][]{ref:24TeDiGa} show that the coldest lines most dominant in the cold component are likely not optically thick, and can therefore further increase in flux with ice enhancement.

\subsection{Gap depth and resolution}
The proposal from the modelling side is that we may conclude how deep the gaps in the outer disc are or where they are located by comparing the inner disc composition to the expected composition from models \citep{ref:24MaSaBi}. This is an interesting idea, and we note that the differences in distances to the objects and therefore the achievable resolution is varying in this sample. Some substructures that appear to be much deeper in one disc compared to the other, may simply be a result of a resolution difference. However, in our discussions above we have shown that there are currently too many uncertainties related to age, formation and transport timescales, volatile desorption, and gap properties as a result of formation mechanism preventing us from doing so with certainty. Instead, the reverse may be of interest, where the inner disc spectrum and visible outer disc structure can inform some of the aforementioned models, to see if the currently observable substructure is deep, shallow, or insufficient to explain the observed inner disc emission. This way it may also be possible to ascertain whether gaps that only block dust pebbles and not small dust grains or gas can even cause sufficiently different inner disc compositions, or whether the current notion of sequential snowlines is realistic. Interesting objects for these kinds of studies would be discs presented here that show some cold \ce{H2O} excess and have deep gaps in millimetre wavelength. For example, Sz~98, which has a deep gap close to the star, but still shows indications for replenishment by \ce{H2O}-ice. Furthermore, SY~Cha has a gap of several tens of astronomical units, along with a cold \ce{H2O}-excess. If transport of pebbles is blocked by these gaps, the influx of ices on small dust grains may also be able to cause this excess.

Furthermore, it it would be useful to identify whether the gaps in the dust continuum in this work are also gaps in the gas. To constrain this, we would need high resolution ALMA observations of \ce{CO} isotopologues, as in \citet{ref:16MaDiBr}. If a gap exists in both the dust and the gas, it is likely that, in addition to pebbles, small dust grains and gas are also not able to reach the inner disc as quickly. In discs where this is the case, the inner disc spectrum may appear more depleted in many commonly detected species and in particular \ce{H2O}, as was discussed above.

\subsection{Implications for planet formation}
It has been posited in the past that the \ce{C/O} ratio of a planet can be traced back to where it formed in the disc, due to natural changes in the \ce{C/O} ratio at different radial positions from the star \citep{ref:11ObMuBe}. However, as was noted previously, this sequence in snowlines may not be entirely representative of reality where volatiles may be trapped in \ce{H2O} ice \citep{ref:04CoAnCh,ref:24LiKiGa}. So far, a planet's atmospheric \ce{C/O} has proven to be a poor tracer, due to phases of migration and other assumptions regarding a planet's formation history and partitioning between the planet envelope and core changing the resulting \ce{C/O} \citep[see e.g.][]{ref:16MoBoMo,ref:22MoMoBi}, and more tracers are now being taken into account \citep[e.g.][]{ref:22PaTuSc}. Furthermore, the atmospheric \ce{C/O} may not be representative for the planet's bulk \ce{C/O}. Nevertheless, the species available at the time of accretion do dictate a planet's resulting composition.

In Sect.~\ref{sec:disks} we note that the presence of substructures within the \ce{CO} and \ce{CH4} snowlines may reduce the gaseous \ce{C/O} ratio in the inner disc, if gas diffusion across the gap is inhibited, such as for photoevaporation. On the other hand, if volatiles are trapped, the presence of a gap outside the \ce{H2O} snowline may prevent replenishment of oxygen in general, and the carbon rich, but overall depleted, gas can still reach the inner disc to increase its C/O. While it is unlikely that planet formation completely cuts off the inner disc from the influence of the outer disc, be it dust or gas transport, if some transport is moderated sufficiently early, planets forming in the inner disc can be more significantly replenished in \ce{H2O}. Whether or not the location of the outer planet matters compared to the different snowlines depends on whether there is actually a sequence in sublimation of ices, or if most of the \ce{CO2}, \ce{CO}, and \ce{CH4} can be trapped in \ce{H2O} ice.

\citet{ref:21MaBoKr} also found that the presence of a gap influences the \ce{C/O} and \ce{C/H} ratios in the outer disc. It is therefore possible that a giant outer planet, massive enough to carve a gap in the disc, could influence the formation conditions in the inner disc. This might bring to mind our Solar System, where Jupiter could have been this giant planet carving a gap \citep{ref:16MoBiCr,ref:19HaWeWi}, and Earth could have been a planet forming in the inner disc influenced by this. In fact, Jupiter has been dated to have formed early in the Solar System's evolution \citep{ref:17KrBuBu}, though this would have reportedly blocked \ce{H2O}-ice and resulted in a ‘dry' inner disc. Instead, we propose that, while eventually the inner disc may dry out, the presence of a gap within the snowlines of the major carbon carriers may rather regulate the amount of incoming dust grains and pebbles, allowing for a somewhat enhanced \ce{H2O} abundance in the inner disc over extended timescales. Therefore, a planet in the outer disc may instead allow for prolonged formation of water-rich worlds in the inner disc, when he outer gas giant forms sufficiently early in the disc lifetime.

\section{Conclusion} \label{sec:conclusion}
In this work, we aim to link the outer dust disc structure to the inner disc gas composition, based on ALMA and MIRI/MRS, respectively. Modelling efforts show clear indications of the influence of substructures on the \ce{H2O} enhancement and volatile C/O ratio, but also on the timescales for which these enhancement scenarios last.

The sample of ten discs presented here shows quite some variation in inner disc emission and outer disc structure. Overall, transport does not seem to be fully prevented. SY~Cha and PDS~70 contain deep, wide gaps that are not fully devoid of gas. This is reflected in their inner disc composition, where SY~Cha especially shows a strong cold \ce{H2O} excess. Similarly, Sz~98 has one of the deepest gaps close to the star in the sample, and also shows a clear cold \ce{H2O} excess. Perhaps the gaps are able to stop the pebbles from reaching the inner disc, but the gas and small dust may still bring in a sufficient abundance of ices.

Furthermore, the discs that do not show a clear excess in a cold \ce{H2O} reservoir show stronger emission from the carbon-bearing species, and vice versa. This is especially clear in \ce{C2H2} and \ce{HCN}, but less so for \ce{CO2}, which is likely influenced by other processes and still contains oxygen.

Two factors that are important to consider in future work are the uncertainties related to age and gap formation timescales and whether volatiles are mostly trapped in \ce{H2O} ice or whether ices are reprocessed into species with higher binding energies. Both dictate whether the gap or its location with respect to the \ce{H2O}, \ce{CO2}, and \ce{CO} snowlines truly have an influence on the inner disc composition.

\begin{acknowledgements}

We thank Andrea Banzatti for useful discussions related to the line selection for the water spectrum, and Feng Long for sharing the DL~Tau continuum measurement set. We also thank the anonymous referee for their thorough and useful feedback. \\

This work is based on observations made with the NASA/ESA/CSA James Webb Space Telescope. The data were obtained from the Mikulski Archive for Space Telescopes at the Space Telescope Science Institute, which is operated by the Association of Universities for Research in Astronomy, Inc., under NASA contract NAS 5-03127 for JWST. These observations are associated with programmes \#1282 and \#1640.

This paper makes use of the following ALMA data: ADS/JAO.ALMA\#2016.1.00484.L, ADS/JAO.ALMA\#2016.1.01164.S, ADS/JAO.ALMA\#2016.1.01370.S, ADS/JAO.ALMA\#2017.1.01167.S, ADS/JAO.ALMA\#2018.1.01458.S, ADS/JAO.ALMA\#2019.1.00607.S. ALMA is a partnership of ESO (representing its member states), NSF (USA) and NINS (Japan), together with NRC (Canada), NSTC and ASIAA (Taiwan), and KASI (Republic of Korea), in cooperation with the Republic of Chile. The Joint ALMA Observatory is operated by ESO, AUI/NRAO and NAOJ. \\
The project leading to this publication has received support from ORP, that is funded by the European Union's Horizon 2020 research and innovation programme under grant agreement No 101004719 [ORP].


A.C.G. acknowledges support from PRIN-MUR 2022 20228JPA3A “The path to star and planet formation in the JWST era (PATH)” funded by NextGeneration EU and by INAF-GoG 2022 “NIR-dark Accretion Outbursts in Massive Young stellar objects (NAOMY)” and Large Grant INAF 2022 “YSOs Outflows, discs and Accretion: towards a global framework for the evolution of planet forming systems (YODA)”.

G.P. gratefully acknowledges support from the Carlsberg Foundation, grant CF23-0481 and from the Max Planck Society.

E.v.D. acknowledges support from the ERC grant 101019751 MOLDISK and the Danish National Research Foundation through the Center of Excellence ``InterCat'' (DNRF150). 

T.H. and K.S. acknowledge support from the European Research Council under the Horizon 2020 Framework Program via the ERC Advanced Grant Origins 83 24 28. 

I.K., A.M.A., and E.v.D. acknowledge support from grant TOP-1 614.001.751 from the Dutch Research Council (NWO).

I.K., J.K., and T.K. acknowledge funding from H2020-MSCA-ITN-2019, grant no. 860470 (CHAMELEON).

B.T. is a Laureate of the Paris Region fellowship program, which is supported by the Ile-de-France Region and has received funding under the Horizon 2020 innovation framework program and Marie Sklodowska-Curie grant agreement No. 945298.

D.G. thanks the Belgian Federal Science Policy Office (BELSPO) for the provision of financial support in the framework of the PRODEX Programme of the European Space Agency (ESA).

D.B. has been funded by Spanish MCIN/AEI/10.13039/501100011033 grants PID2019-107061GB-C61 and No. MDM-2017-0737. 

M.T. and M.V. acknowledge support from the ERC grant 101019751 MOLDISK.
\end{acknowledgements}

\bibliography{bib.bib} 
\bibliographystyle{aa.bst} 

\begin{appendix}

\section{Selected H$_2$O lines and line fluxes}

\begin{figure*}[t]
    \centering
    \includegraphics[width=\textwidth]{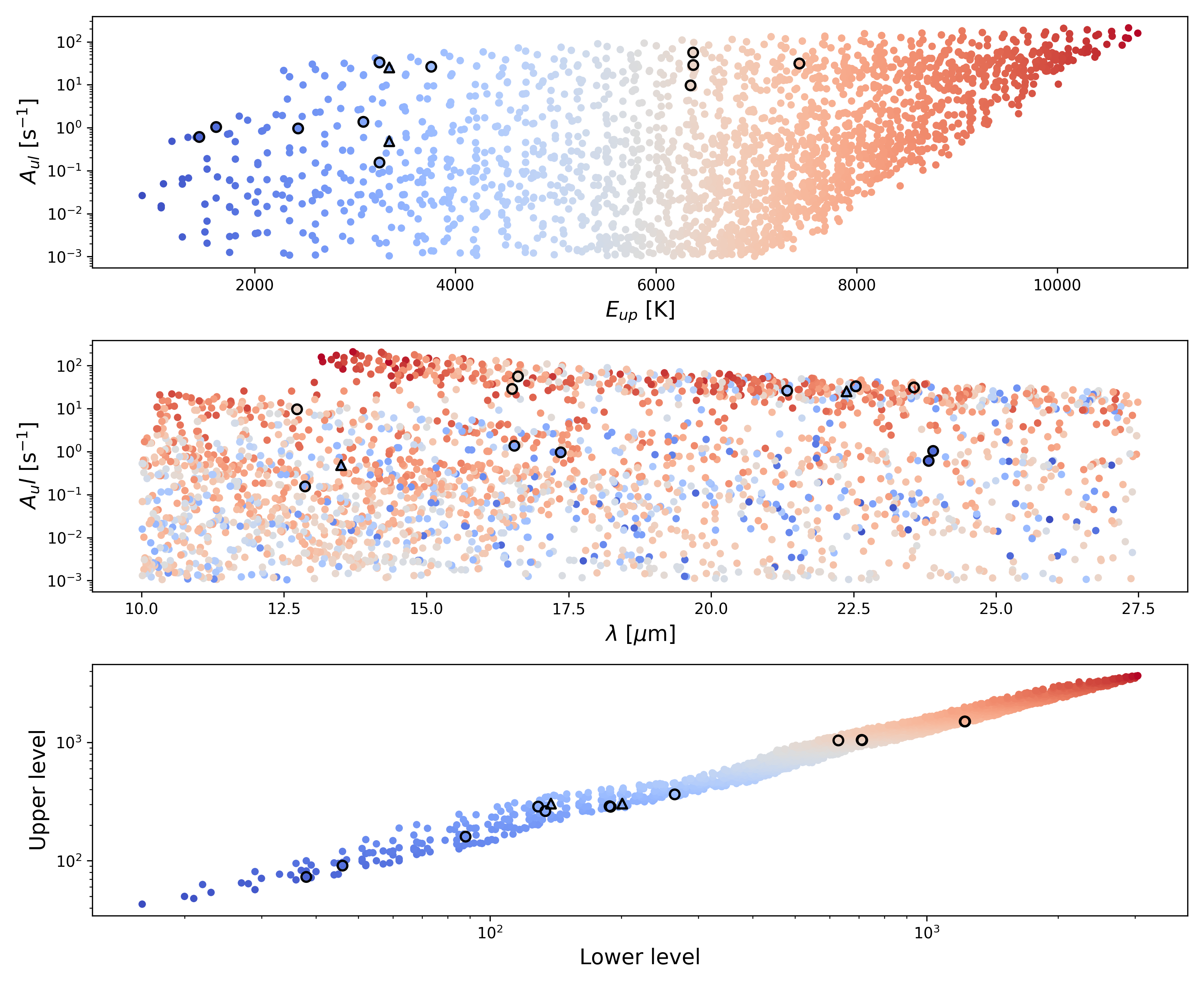}
    \caption{Lines used to calculate the warm and cold water line fluxes. The open circles are single lines. The open triangles are a line pair with the same upper level used to probe the column density in the cold water gas.}
    \label{fig:lines_used}
\end{figure*}

In Fig.~\ref{fig:lines_used} and Table~\ref{tab:lines}, we summarise the \ce{H2O} lines that are considered in this study.

\begin{table*}[h!]
\caption{Summary of \ce{H2O} lines with quantum numbers 000--000 ($v_1v_2v_3$, upper-lower levels) used in this work.}
\centering
\label{tab:lines}
\begin{tabular}{llll} \hline \hline
\begin{tabular}[c]{@{}l@{}}Wavelength\\ {[}$\mu$m{]}\end{tabular} & \begin{tabular}[c]{@{}l@{}}Transition (upper-lower levels)\\ Level format: $J_{K_{a}K_{c}}$\end{tabular} & \begin{tabular}[c]{@{}l@{}}$A_{ul}$\\ {[}s$^{-1}${]}\end{tabular} & \begin{tabular}[c]{@{}l@{}}$E_u$\\ {[}$K${]}\end{tabular} \\ \hline
\multicolumn{4}{c}{Warm}                                                                                                                                                                                                                                                                                                 \\ \hline
23.560                                                            & 21$_{\text{ }2\text{ }19}$--20$_{\text{ }2\text{ }18}$                                                                        & 31.26                                                             & 7428.48                                                   \\
23.559                                                            & 21$_{\text{ }3\text{ }19}$--20$_{\text{ }3\text{ }18}$                                                                        & 31.26                                                             & 7428.49                                                   \\
16.608                                                            & 16$_{\text{ }9\text{ }8}$--15$_{\text{ }8\text{ }7}$                                                                        & 56.31                                                             & 6369.60                                                   \\
16.505                                                            & 17$_{\text{ }7\text{ }10}$--16$_{\text{ }6\text{ }11}$                                                                      & 28.79                                                             & 6371.03                                                   \\
12.729                                                            & 17$_{\text{ }7\text{ }11}$--16$_{\text{ }4\text{ }12}$                                                                      & 9.669                                                             & 6344.03                                                   \\ \hline
\multicolumn{4}{c}{Cold}                                                                                                                                                                                                                                                                                                 \\ \hline
23.896                                                            & 8$_{\text{ }4\text{ }5}$--7$_{\text{ }1\text{ }6}$                                                                          & 1.038                                                             & 1615.32                                                   \\
23.817                                                            & 8$_{\text{ }3\text{ }6}$--7$_{\text{ }0\text{ }7}$                                                                          & 0.609                                                             & 1447.58                                                   \\
22.538                                                            & 10$_{\text{ }8\text{ }3}$--9$_{\text{ }7\text{ }2}$                                                                         & 33.15                                                             & 3243.29                                                   \\
22.538                                                            & 10$_{\text{ }8\text{ }2}$--9$_{\text{ }7\text{ }3}$                                                                         & 33.17                                                             & 3243.29                                                   \\
21.333                                                            & 12$_{\text{ }7\text{ }6}$--11$_{\text{ }6\text{ }5}$                                                                        & 26.27                                                             & 3759.21                                                   \\
17.358                                                            & 11$_{\text{ }2\text{ }9}$--10$_{\text{ }1\text{ }10}$                                                                       & 0.9617                                                            & 2432.47                                                   \\
16.544                                                            & 11$_{\text{ }6\text{ }6}$--10$_{\text{ }3\text{ }7}$                                                                        & 1.37                                                              & 3082.70                                                   \\
12.870                                                            & 10$_{\text{ }8\text{ }2}$--9$_{\text{ }5\text{ }5}$                                                                         & 0.1549                                                            & 3243.39                                                   \\ \hline
\multicolumn{4}{c}{Column density line pair}                                                                                                                                                                                                                                                                             \\ \hline
13.503                                                            & 11$_{\text{ }7\text{ }4}$--10$_{\text{ }4\text{ }7}$                                                                        & 0.4852                                                            & 3340.68                                                   \\
22.375                                                            & 11$_{\text{ }7\text{ }4}$--10$_{\text{ }6\text{ }5}$                                                                        & 25.29                                                             & 3340.68                                                   \\ \hline
\end{tabular}
\end{table*}

\begin{table*}[t]
\caption{Line fluxes of the \ce{H2O} lines discussed in Sect.~\ref{sec:line_int}.}
\label{tab:line_fluxes_h2o}
\centering
\resizebox{1\linewidth}{!}{%
\begin{tabular}{llllllllllllll}
\hline \hline
\textbf{}   & \multicolumn{13}{c}{\begin{tabular}[c]{@{}c@{}}H2O \end{tabular}}                                                                                                                               \\
\textbf{ID} & 17$_{\text{ }7\text{ }11}$--16$_{\text{ }4\text{ }12}$ & 17$_{\text{ }7\text{ }10}$--16$_{\text{ }6\text{ }11}$  & 16$_{\text{ }9\text{ }8}$--15$_{\text{ }8\text{ }7}$ & \multicolumn{1}{c}{\begin{tabular}[c]{@{}c@{}}
21$_{\text{ }2\text{ }19}$--20$_{\text{ }2\text{ }18}$\\ 21$_{\text{ }3\text{ }19}$--20$_{\text{ }3\text{ }18}$ \end{tabular}} & 10$_{\text{ }8\text{ }2}$-90$_{\text{ }5\text{ }5}$ & 11$_{\text{ }6\text{ }6}$--10$_{\text{ }3\text{ }7}$  & 11$_{\text{ }2\text{ }9}$--10$_{\text{ }1\text{ }10}$ & 12$_{\text{ }7\text{ }6}$--11$_{\text{ }6\text{ }5}$ & \multicolumn{1}{c}{\begin{tabular}[c]{@{}c@{}} 10$_{\text{ }8\text{ }2}$--9$_{\text{ }7\text{ }3}$ \\ 10$_{\text{ }8\text{ }3}$--9$_{\text{ }7\text{ }2}$  \end{tabular}} & 8$_{\text{ }3\text{ }6}$--7$_{\text{ }0\text{ }7}$  & 8$_{\text{ }4\text{ }5}$--7$_{\text{ }1\text{ }6}$ & 11$_{\text{ }7\text{ }4}$--10$_{\text{ }4\text{ }7}$ & 11$_{\text{ }7\text{ }4}$--10$_{\text{ }6\text{ }5}$  \\ \hline
GW Lup      & 0.57$\pm$0.19 & 1.02$\pm$0.14 & 0.62$\pm$0.09 & 0.64$\pm$0.13 & 0.46$\pm$0.07 & 0.57$\pm$0.09 & 1.20$\pm$0.12 & 0.76$\pm$0.10 & 1.50$\pm$0.12 & 1.28$\pm$0.10 & 1.23$\pm$0.10 & 1.09$\pm$0.17 & 0.89$\pm$0.12  \\
IQ Tau      & 1.75$\pm$0.27 & 2.61$\pm$0.20 & 2.68$\pm$0.13 & 1.93$\pm$0.19 & 1.00$\pm$0.11 & 1.30$\pm$0.13 & 4.30$\pm$0.18 & 2.04$\pm$0.14 & 2.74$\pm$0.17 & 4.60$\pm$0.15 & 3.91$\pm$0.15 & 2.23$\pm$0.24 & 2.46$\pm$0.17  \\
BP Tau      & 0.76$\pm$0.29 & 2.63$\pm$0.21 & 1.96$\pm$0.14 & 2.63$\pm$0.21 & 0.92$\pm$0.11 & 1.30$\pm$0.14 & 8.32$\pm$0.19 & 4.74$\pm$0.15 & 7.47$\pm$0.18 & 9.00$\pm$0.16 & 9.31$\pm$0.16 & 2.44$\pm$0.26 & 5.63$\pm$0.18  \\
Sz 98       & 2.53$\pm$0.45 & 5.05$\pm$0.32 & 3.14$\pm$0.21 & 3.50$\pm$0.31 & 1.99$\pm$0.18 & 2.07$\pm$0.21 & 12.8$\pm$0.29 & 4.02$\pm$0.23 & 7.42$\pm$0.28 & 19.9$\pm$0.25 & 15.6$\pm$0.24 & 6.41$\pm$0.40 & 6.74$\pm$0.28  \\
CI Tau      & 4.59$\pm$0.44 & 8.13$\pm$0.32 & 6.18$\pm$0.21 & 5.09$\pm$0.31 & 2.17$\pm$0.17 & 3.04$\pm$0.21 & 8.28$\pm$0.29 & 5.54$\pm$0.23 & 8.11$\pm$0.27 & 6.62$\pm$0.24 & 7.39$\pm$0.24 & 5.81$\pm$0.40 & 6.55$\pm$0.28  \\
V1094 Sco   & 0.44$\pm$0.10 & 0.42$\pm$0.07 & 0.19$\pm$0.04 & 0.33$\pm$0.07 & 0.21$\pm$0.04 & 0.25$\pm$0.05 & 0.82$\pm$0.06 & 0.46$\pm$0.05 & 0.89$\pm$0.06 & 0.95$\pm$0.05 & 1.04$\pm$0.05 & 0.57$\pm$0.08 & 0.57$\pm$0.06   \\
DR Tau      & 12.9$\pm$3.70  & 29.2$\pm$2.64 & 23.0$\pm$1.74 & 21.8$\pm$2.59 & 10.3$\pm$1.45 & 17.4$\pm$1.75 & 72.0$\pm$2.39 & 35.9$\pm$1.90 & 54.7$\pm$2.27 & 70.3$\pm$2.03 & 69.2$\pm$2.02 & 28.9$\pm$3.29 & 40.8$\pm$2.30  \\
DL Tau      & 3.13$\pm$0.38 & 3.74$\pm$0.27 & 3.89$\pm$0.18 & 2.37$\pm$0.26 & 1.49$\pm$0.15 & 1.95$\pm$0.18 & 4.99$\pm$0.24 & 2.09$\pm$0.19 & 4.45$\pm$0.23 & 4.55$\pm$0.21 & 5.66$\pm$0.21 & 4.91$\pm$0.34 & 4.22$\pm$0.23  \\
PDS 70      & 0.04$\pm$0.04 & 0.10$\pm$0.03 & 0.06$\pm$0.02 & 0.06$\pm$0.03 & 0.05$\pm$0.02 & 0.05$\pm$0.02 & 0.33$\pm$0.03 & 0.10$\pm$0.02 & 0.28$\pm$0.02 & 0.33$\pm$0.02 & 0.46$\pm$0.02 & 0.24$\pm$0.04 & 0.20$\pm$0.02   \\
SY Cha      & 0.95$\pm$0.27 & 1.59$\pm$0.19  & 0.93$\pm$0.13 & 1.34$\pm$0.19 & 0.70$\pm$0.1   & 0.84$\pm$0.13 & 4.21$\pm$0.17 & 1.42$\pm$0.14 & 2.70$\pm$0.16 & 8.56$\pm$0.15 & 6.05$\pm$0.15 & 1.95$\pm$0.24 & 2.22$\pm$0.17  \\ \hline
\end{tabular}%
}
{\raggedright \textbf{Notes.}  All values are normalised to 140~pc and in 10$^{-15}$ erg s$^{-1}$ cm$^{-2}$. The two columns denoting two lines are actually a combined flux measurement of two blended lines of the same energy, see also Table~\ref{tab:lines}. \par}

\end{table*}

\begin{table*}[t]
\caption{Line fluxes of the carbon-bearing species discussed in Sect.~\ref{sec:line_int}.}
\label{tab:line_fluxes_carbon}
\centering
\begin{tabular}{llll}
\hline \hline
\textbf{ID}   &  \multicolumn{1}{c}{\begin{tabular}[c]{@{}c@{}}CO2 \end{tabular}} & \multicolumn{1}{c}{\begin{tabular}[c]{@{}c@{}}HCN \end{tabular}} & \multicolumn{1}{c}{\begin{tabular}[c]{@{}c@{}}C2H2 \end{tabular}} \\ \hline
GW Lup      &  14.5$\pm$0.88                                                                                             & 11.3$\pm$1.61                                                                                             & 4.45$\pm$1.85                                                                                              \\
IQ Tau      &  3.52$\pm$1.27                                                                                             & 22.9$\pm$2.32                                                                                             & 4.30$\pm$2.66                                                                                              \\
BP Tau      &  5.47$\pm$1.36                                                                                             & 10.3$\pm$2.48                                                                                             & 4.23$\pm$2.84                                                                                              \\
Sz 98       &  9.59$\pm$2.08                                                                                             & 27.6$\pm$3.80                                                                                             & 4.68$\pm$4.36                                                                                              \\
CI Tau      &  18.9$\pm$2.06                                                                                             & 53.2$\pm$3.76                                                                                             & 49.3$\pm$4.31                                                                                              \\
V1094 Sco   & 3.93$\pm$0.44                                                                                             & 9.21$\pm$0.81                                                                                             & 11.6$\pm$0.92 \\
DR Tau     & 64.8$\pm$17.1                                                                                             & 211$\pm$31.3                                                                                              & 95.0$\pm$35.9                                                                                              \\
DL Tau     & 8.67$\pm$1.75                                                                                             & 30.3$\pm$3.19                                                                                             & 93.2$\pm$3.66                                                                                             \\
PDS 70      & 0.83$\pm$0.18                                                                                             & 0.70$\pm$0.34                                                                                             & 0.76$\pm$0.39                                                                                              \\
SY Cha      & 6.12$\pm$1.24                                                                                             & 11.8$\pm$2.27                                                                                             & 3.69$\pm$2.60                                                                                               \\ \hline
\end{tabular}

{\raggedright \textbf{Notes.}   All values are normalised to 140~pc and in 10$^{-15}$ erg s$^{-1}$ cm$^{-2}$. \par}

\end{table*}

\section{Extended discussion of scenarios per disc}
\label{app:extended_discussion}
\subsection{DR Tau}
We start this discussion with DR~Tau, the most \ce{H2O}-rich source in this sample by far, but only a marginal excess of cold \ce{H2O} \citep{ref:24TeDiGa}. Its age is estimated to be around 1.8 to 5.9~Myrs (see Table~\ref{tab:sample}). It is certainly not the weakest in its emission strength of the carbon species, but not the strongest either. This likely places it in one of the \ce{H2O}-rich scenarios of Fig.~\ref{fig:scenarios}. Additionally, in Fig.~\ref{fig:radial_profiles} it can be seen that its dust disc is relatively small, with some weaker substructure at the current spatial resolution, with the \ce{CO} and \ce{CH4} snowlines clearly outside of the inner-most substructure. Assuming its age is a few million years (see Table~\ref{tab:sample}), these substructures are likely not completely blocking the incoming dust pebbles. If they were, the \ce{H2O} in the inner disc might already have been depleted on these timescales. This leaves the possibilities of the shallow and moderate gaps (cases 1 and 2, respectively) to be more likely, with scenario 2 the most likely if the disc is on the older end of the age estimates.

\subsection{BP Tau}
As is mentioned in Table~\ref{tab:sample}, the age of BP~Tau is estimated to be in the range of 1.0 to 3.4~Myrs. The substructure identified in BP~Tau is either a very close-in gap, or a cavity of a few astronomical units. Furthermore, in Table~\ref{tab:emission_summary} it can be seen that it is likely somewhat depleted in carbon, but similar in oxygen to DR~Tau. It could be that BP~Tau is younger, resulting in more of its carbon-rich ices and gases still making their way inwards from the outer disc. Its \ce{H2O}$_\text{warm}$ line fluxes are weaker compared to its \ce{H2O}$_\text{cold}$. According to \citet{ref:17BaPoSa}, the presence of some gas depletion in the inner disc could result in weaker warmer \ce{H2O}. Rather than from the outside in, like dust migration inwards would cause, the \ce{H2O} gas is depleted from the inside out. A similar effect is modelled in \citet{ref:24VlDiTa}, where for cavity sizes of a few astronomical units like in BP~Tau, first the \ce{H2O} flux is increased compared to \ce{CO2}. Initially, the snowlines are simply shifted outwards, increasing the flux of the molecular lines. Additionally, the accretion luminosity of BP~Tau indicates that, while it is on the lower end in the sample, material is still being accreted onto the star, meaning the cavity is not completely void of material. \citet{ref:24VlDiTa} show that in this case, the increase in relative \ce{CO2} flux becomes less prominent as well. Due to the potential presence of the inner cavity of a few astronomical units, BP~Tau may not fall under any of the categories shown in Fig.~\ref{fig:scenarios}, although its evolution may be similar to scenario 1, depending on how the cavity evolves.

\subsection{Sz 98}
The age of Sz~98 has quite a wide estimated range (as is mentioned in Table~\ref{tab:sample}), but we assume it is similar to the other discs in this sample; of the order of a few million years. As is presented in Table~\ref{tab:emission_summary}, Sz~98 may be somewhat enhanced in oxygen compared to DR~Tau. It is bright in \ce{H2O} and may have a somewhat higher column density of \ce{CO2} than DR~Tau, but it appears weaker in the other carbon bearing species, especially \ce{C2H2}. Additionally, based on its profile in Fig.~\ref{fig:radial_profiles}, we observe that its disc contains a strong substructure near the star, inside the snowlines of \ce{CO} and \ce{CH4}. Similarly to DR~Tau, if it is of the order of a few million years old, this inner gap should result in relatively quick depletion of \ce{H2O} if it were blocking all the dust. Since this is not the case, we might expect it to fall under the shallow and moderate gap categories, like DR~Tau. However, in this scenario the carbon-rich gases from the outer disc could still freely move towards the inner disc. Since the spectrum of the inner disc presents itself as relatively carbon-poor, either this has not yet happened, or the gas is also blocked from reaching the inner disc. In this case, the gap may instead be caused by photoevaporation, rather than processes akin to planet formation. In case 6 or 7 of Fig.~\ref{fig:scenarios}, the inner disc quickly becomes \ce{H2O} rich as the carbon grains and gases are completely blocked from reaching the inner disc. In addition, the gas in the inner disc is spread back outward across the \ce{H2O} snowline, and \ce{H2O} can migrate back as ice across its snowline. This results in a long \ce{H2O} equilibrium cycle \citep{ref:24LiBiHe}, where the inner disc becomes \ce{H2O} rich and carbon poor, and we can observe a cold \ce{H2O} reservoir for extended periods of time. However, this requires a low viscosity to maintain the inner disc for an extended period of time.

\subsection{IQ Tau}
Interestingly, the outer disc structure of IQ~Tau appears quite similar to that of DR~Tau (see Fig.~\ref{fig:radial_profiles}). The age estimates are also not dissimilar, with estimates ranging from 2.2 to 8.3~Myrs (see Table~\ref{tab:sample}). It shows some of the stronger \ce{H2O} line fluxes in the sample, especially in terms of its warmer reservoir. In Sect.~\ref{sec:results} we observed that its colder \ce{H2O} reservoir is weaker than DR~Tau in terms of emission strength, but the relative strength of the first and third line in the 23.8-24~$\mu$m quadruplet indicates that some excess \ce{H2O}$_\text{cold}$ may be present, albeit on the weaker end in the sample. Furthermore, its \ce{C2H2} and \ce{CO2} $Q$-branch emission appears weaker than DR~Tau. The excess in the range of the hot branch of \ce{CO2} may point towards a higher column density, but due to the relatively weak \ce{CO2} $Q$-branch it is more likely that this is more strengthened by an \ce{H2O} line in this range. Its luminosity is lower than DR~Tau, which may have two consequences. First, the temperatures may be lower in the inner disc, resulting in \ce{CO2} formation being favoured over \ce{H2O} formation, and therefore in a higher column density for the former \citep[e.g.][]{ref:09WoThKa,ref:09GlMeNa,ref:13DiHeNe}. Due to the weak $Q$-branch, this may be less realistic. Second, the distance between the star and the \ce{H2O} snowline reduces, in turn reducing the possible emitting area for the colder \ce{H2O} reservoir. Based on the estimated snowline location in Fig.~\ref{fig:radial_profiles}, it could be that the emitting area of \ce{H2O}, and especially the \ce{H2O}$_\text{cold}$, may simply be less than in DR~Tau. Furthermore, due to this lower luminosity, the \ce{CO} and \ce{CH4} snowlines are within the detected substructures in IQ~Tau. These substructures may be leaky to dust as suspected for DR~Tau, and IQ~Tau may be moving from the younger versions of scenarios 1 or 2, to a more processed, more \ce{CO2} rich disc as is shown in Fig.~\ref{fig:scenarios}, and modelled in \citet{ref:24SeVlDi}.

\subsection{CI Tau}
The disc structure shown in Fig.~\ref{fig:radial_profiles} of CI~Tau shows similarities to Sz~98, with a deep gap relatively close to the star. Its age is estimated to be on the 1.4 to 4.5~Myr range (Table~\ref{tab:sample}). Contrary to Sz~98, however, it is stronger in \ce{C2H2}, much less so in \ce{H2O}$_\text{cold}$, and similar to DR~Tau in all other examined emission (see Table~\ref{tab:emission_summary}). While the colder \ce{H2O} reservoir may be smaller, its warmer \ce{H2O} is still relatively strong in the sample. Therefore, CI~Tau may be in the process of accreting its \ce{H2O} gas and relatively little may be coming in. In that case, the detected most dominant \ce{H2O} emission is from a warmer reservoir closer to the star, where the outer regions closer to the snowline are slowly depleted of \ce{H2O}. This would indicate that its inner gap is relatively deep, corresponding to scenario 4 in Fig.~\ref{fig:scenarios}, allowing this phase to start earlier than for DR~Tau, whose gap may be leakier. Similarly to Sz~98, the \ce{CO} and \ce{CH4} snowlines are clearly outside of the most prominent substructure, yet its inner disc does not seem to be significantly depleted in carbon. This, along with the lack of a strong cold \ce{H2O} signature, would exclude photoevaporation gaps, as opposed to Sz~98. The only other significant difference with Sz~98 is the stellar luminosity, which is significantly lower than Sz~98.

\subsection{GW Lup}
As is mentioned in Table~\ref{tab:sample}, the age of GW~Lup is estimated to be from 0.8 to 5.0~Myrs. GW~Lup is extremely strong in \ce{CO2}, high in the sample in its warmer \ce{H2O} emission, but seemingly depleted in cold \ce{H2O}. Furthermore, its strongest substructure appears relatively far from the star, around 74~au (see Fig.~\ref{fig:radial_profiles} and App.~\ref{app:ALMA}), several very weak substructures in the radial profile appear closer to the star, but these are by far the weakest variations in the sample. The strongest gap may be too far out to block a significant part of the \ce{H2O} ices from reaching the inner disc \citep{ref:21KaPiKr,ref:23KaPiKr}. This would point to scenarios 3 and 5 in Fig.~\ref{fig:scenarios} being more likely. The \ce{H2O}$_\text{warm}$ is stronger than \ce{H2O}$_\text{cold}$, indicating the \ce{H2O} reservoir is no longer strongly being replenished, and in the process of accreting onto the star. In turn, the disc is in the \ce{CO2} enhanced phase \citep{ref:24SeVlDi}, as is indicated by the strong \ce{CO2} emission and the detection of $^{13}$CO$_2$. Therefore, more evolved cases of scenarios 3 and 5 are likely. Moreover, the luminosity of GW~Lup is relatively low in this sample, and may therefore, similarly to IQ~Tau, have its snowlines closer to the star, allowing them to reach a more evolved stage more rapidly.

\subsection{DL Tau}
The age of DL~Tau is estimated to fall between 1.9 to 6.3~Myrs (Table~\ref{tab:sample}). DL~Tau is one of the weakest \ce{H2O} emitters out of the full discs presented in this sample, and does not appear significantly stronger in its \ce{CO2} emission. It is therefore unlikely that a significant portion of its oxygen is locked up in \ce{CO2} rather than \ce{H2O}. Its \ce{HCN} and \ce{C2H2} emission on the other hand, we show to be comparatively much stronger, as summarised in Table~\ref{tab:emission_summary}. The inner disc therefore appears relatively depleted in oxygen, and enhanced in carbon. The relatively warm remaining \ce{H2O} reservoir indicates the emission may originate from closer to the star, and not out to the snowline. Its radial profile in Fig.~\ref{fig:radial_profiles} shows a range of relatively shallow substructures, with one deeper gap present around 60~au. Similarly to GW~Lup, this gap farther out may not be as effective at blocking \ce{H2O} ice, allowing for evolution closer to scenarios 1, 3 or 5, in a more evolved stage of its lifetime. Alternatively, the gap closer to the star could be very deep, in which case the disc can be younger, as its \ce{H2O} rich phase would pass after less than a million years \citep{ref:24MaSaBi}.

\subsection{V1094 Sco}
The final full disc, V1094~Sco, has even larger uncertainties in its age compared to Sz~98 (see Table~\ref{tab:sample} and the related note). It has the lowest \ce{H2O} flux overall in this sample, and it likely does not contain an exceptionally strong cold \ce{H2O} reservoir. It is much stronger in \ce{C2H2}, \ce{HCN}, and \ce{CO2}, as is shown in Table~\ref{tab:Tcond}. However, its \ce{CO2} is not as strong as in GW~Lup, which, along with the strength of the other carbon species, indicates that the inner disc might have a relatively high C/O. Taking a look at its outer disc structure then, in Fig.~\ref{fig:radial_profiles}, it can be seen that the most prominent substructure is located relatively close to the star, around 17~au. This gap may be deep, resulting in an evolved scenario 4, where the \ce{H2O} enhanced phase very rapidly passes and the inner disc quickly presents itself as more carbon-rich. Since the \ce{CO} and \ce{CH4} snowlines are present outside this inner substructure, it is likely that the carbon gases are able to freely migrate across the gap. Alternatively, if the gap is shallow (scenario 1) or moderate (scenario 2), the disc must be older \citep{ref:24MaSaBi}. For the shallow case this means likely older than a few million years, which seems feasible, but for the moderate gap this may mean an age of well over 10~Myrs, which may be less realistic.

\subsection{SY Cha}
The estimated age of SY~Cha is 3~Myrs, as was previously mentioned in Table~\ref{tab:sample}). Out of the two discs with extremely wide gaps, SY~Cha clearly has the strongest in \ce{H2O} emission, and indeed all molecular emission overall, as is also discussed in \citet{ref:24ScHeCh}. In fact, while its warm \ce{H2O} lines are not always confidently detected, the strength of the detected cold \ce{H2O} lines is the highest overall. Along with Sz~98, it is the only source with indications for a more enhanced cold \ce{H2O} reservoir. In all other species, it is similar to DR~Tau as well. The inner disc of SY~Cha is small and may itself have a smaller cavity near the star, whereas the deep cavity from the inner disc up to $\sim$70~au is likely not fully depleted of gas \citep{ref:23OrMoMu}. The smaller \ce{H2O}$_\text{warm}$ flux may in this case have similar origins to BP~Tau. SY~Cha may be somewhat depleted in material close to the star, reducing the flux of the shorter wavelength, warmer lines \citep{ref:17BaPoSa,ref:20BaPaBo}. Additionally, the detection of gas within the deep outer cavity indicates that it is not stopping all material from flowing inwards. Especially carbon dominated gases can find their way inwards from the outer disc, along with some smaller dust grains coupled to the gas, which may be able to replenish some of the inner \ce{H2O} reservoir, as is also proposed by \citet{ref:24ScHeCh}.

\subsection{PDS 70}
The story may be somewhat similar for PDS~70 to that of SY~Cha, although PDS~70 is estimated to be slightly older than SY~Cha (4.4 to 6.4~Myrs, versus 3~Myrs, Table~\ref{tab:sample}). It is also a disc with a major gap of tens of astronomical units wide, and a relatively weak inner disc with a radius of 18~au \citep{ref:21BeBaFa}. The disc MRS spectrum, published in \citet{ref:23PeChHe}, shows some \ce{H2O} emission, but very little, if any \ce{C2H2} and \ce{HCN}, and perhaps an elevated \ce{CO2} flux compared to DR~Tau. Again, the presence of the wide outer gap does not prevent the oxygen species from being more dominant in the inner disc, as is discussed in \citet{ref:23PeChHe}. Furthermore, \citet{ref:24PiBeWa} and \citet{ref:24JaWaKa} theorise that aside from some gas, small dust may also be moving from the outer disc to the inner disc, perhaps replenishing some of the inner disc's oxygen. This is also in line with the observations in \citet{ref:23PeChHe}. In the case of PDS~70, the large gap may be more depleted and its oxygen replenishment may be less significant than for SY~Cha, resulting in a more \ce{CO2} dominant disc, or it is simply older, similarly to some of the scenarios for the full discs as is shown in Fig.~\ref{fig:scenarios}.

\section{Estimating radii and gaps from ALMA}
\label{app:ALMA}

\subsection{ALMA observations}

\begin{figure*}[ht!]
    \centering
    \includegraphics[width=\textwidth]{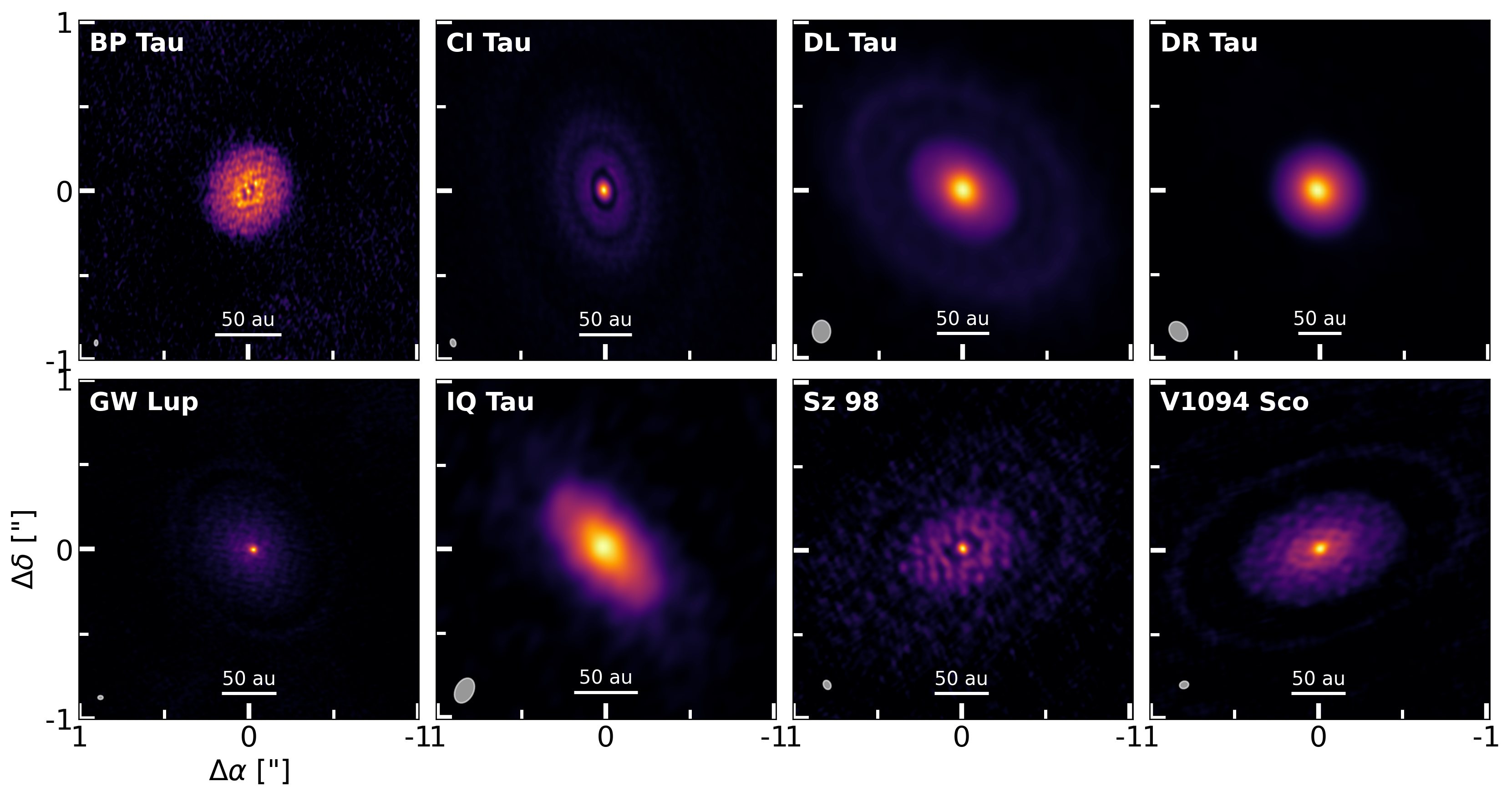}
    \caption{ALMA continuum images of the various sources. The resolving beams are displayed in the lower left corner and the horizontal bars indicate scales of 50 au.}
    \label{fig:Continuum}
\end{figure*}

As we are interested in the substructures closest to the host star, we have retrieved all the highest available spatial resolution ALMA observations publicly available for our sources. We list the distances, inclinations, position angles, and some of the observation properties in Table~\ref{tab:Sample}. As for many of the sources, except BP~Tau, the inclination and the position angle are well constrained, we use the literature values while fitting the visibilities. For BP~Tau, we estimated updated values using the CASA-task \textsc{imfit} by fitting an elliptical Gaussian. Additionally, we used \textsc{imfit} to estimate the phase offsets ($\Delta\alpha$ and $\Delta\delta$, both in arcsec) for each source and corrected for these offsets using the CASA-task \textsc{fixvis}. For some sources (CI~Tau and GW~Lup), additional shifts were determined from the image plane, ensuring that the peak flux in the model falls on the same position as the peak flux in the image. For only three of our sources (CI~Tau, DR~Tau, and IQ~Tau), self-calibration yielded an increased signal-to-noise ratios (S/N). Therefore, we performed two rounds of phase-only self-calibration (solution intervals of `inf' and `30s') for these sources. We have imaged (down to a threshold of 1.5$\times$ the root-mean square) the dust continuum of all our sources using the `Briggs' weighting scheme with a robust parameter of +0.5. Fig.~\ref{fig:Continuum} displays the continuum images and the resolving beams are highlighted in Table~\ref{tab:Sample}.
\begin{table*}[ht!]
    \centering
    \caption{Source properties and band 6 ALMA observations of the eight T~Tauri discs.}
    \resizebox{1\linewidth}{!}{%
    \begin{tabular}{c c c c c c c c c}
        \hline\hline
        Source & Inc. & PA & ALMA PID (PI) & Frequency & Reference & Reference & Beam & Ref. \\
        & [\degree] & [\degree] & & [GHz] & frequency & wavelength & ["$\times$" (\degree)] & \\
        &  &  & &  & [GHz] & [mm] &  & \\
        \hline
        BP~Tau & 23.6 & 163.9 & 2019.1.00607.S (F. Long) & 215-234 & 230.7 & 1.3 & 0.034$\times$0.020 (-2.5) & This work \\
        CI~Tau & 50.0 & 11.2 & 2016.1.01370.S (C. Clarke) & 223-243 & 224.0 & 1.3 & 0.046$\times$0.032 (15.9) & \citet{ref:18LoPiHe} \\
        DL~Tau & 45.0 & 52.1 & 2016.1.01164.S (G. Herczeg) & 217-234 & 218.0 & 1.4 & 0.133$\times$0.108 (-2.3) & \citet{ref:18LoPiHe} \\
        DR~Tau & 5.4 & 3.4 & 2016.1.01164.S (G. Herczeg) & 217-234 & 218.0 & 1.4 & 0.128$\times$0.100 (38.3) & \citet{ref:19LoHeGr} \\
        GW~Lup & 38.7 & 37.7 & 2016.1.00484.L (S. Andrews) & 230-248 & 232.6 & 1.3 & 0.030$\times$0.023 (-89.2) & \citet{ref:18HuAnDu} \\
        IQ~Tau & 62.1 & 42.4 & 2016.1.01164.S (G. Herczeg) & 217-234 & 218.0 & 1.4 & 0.157$\times$0.106 (-28.0) & \citet{ref:18LoPiHe} \\
        Sz~98 & 47.1 & 111.6 & 2018.1.01458.S (Yen, H-W) & 217-234 & 233.0 & 1.3 & 0.056$\times$0.043 (27.3) & \citet{ref:17TaTeNa} \\
        V1094~Sco & 53.0 & 109.0 & 2017.1.01167.S (S. Perez) & 229-248 & 232.5 & 1.3 & 0.054$\times$0.043 (-73.9) & \citet{ref:18TeDiAn} \\
        \hline
    \end{tabular}%
    }
    \label{tab:Sample}
\end{table*}

\subsection{Visibility fitting}
Given the non-uniformity of the ALMA sample, we have fitted the visibilities to infer potentially `hidden' substructures in the discs that cannot be retrieved from the cleaned images. We have followed the approach by \citet{ref:16ZhBeBl}, where a Hankel transform acts as the link between the deprojected $uv$-distance and the radial brightness distribution ($I(\theta)$) \citep{ref:99Pe}:
\begin{align}
    u' & = \left(u\cos(\phi) - v\sin(\phi)\right) \times \cos(i), \\
    v' & = u\sin(\phi) + v\cos(\phi), \\
    V(\rho) & = 2\pi\int^\infty_0 I_\nu(\theta)\theta J_0(2\pi\theta\rho)\textnormal{d}\theta.
\end{align}
Here, $i$ and $\phi$ are the disc's inclination and position, $\rho=\sqrt{u'^2+v'^2}$ denotes the deprojected $uv$-distance (given in units of $\lambda$, and $\theta$ is the radial angular scale as seen from the disc's centre. Finally, $J_0$ denotes the zeroth-order Bessel function of the first kinds. The model brightness distribution consists of a set of Gaussian functions, inspired by the peaks seen in the visibilities and modulated by a sinusoidal function that has a spatial frequency of $\rho_i$,
\begin{align}
    I(\theta) = & \frac{a_0}{\sqrt{2\pi}\sigma_0}\exp\left(\frac{\theta^2}{-2\sigma_0^2}\right) +  \nonumber \\
              & \sum_i \cos\left(2\pi\theta\rho_i\right) \times \frac{a_i}{\sqrt{2\pi}\sigma_i}\exp\left(-\frac{\theta^2}{2\sigma_i^2}\right).
              \label{eq:gauss}
\end{align}
This function yields the \{$a_0$, $\sigma_0$, $a_i$, $\sigma_i$, $\rho_i$\} as the set of free parameters. For our sample, we report the minimum of additional Gaussian functions that results in a good fit to all the visible features in the visibilities.

We have deprojected the visibilities using the inclinations and position angles listed in Table \ref{tab:Sample}. To ensure that we are able to fit the visibilities properly, we have visually compared the visibilities of all spectral windows (line and continuum) with those of the continuum spectral windows alone. For the line spectral windows, we masked visible lines, ensuring that they only contain continuum emission. In the cases of DR~Tau, GW~Lup, and Sz~98, we found that the uncertainties of the visibilities of the continuum spectral windows were significantly lower than those when including all spectral windows. Therefore, we only used the continuum spectral windows for those three sources. In addition, we only use the continuum spectral windows for DL~Tau, as the measurement set was obtained through private communication. For all other sources we have used all spectral windows.

We bin the visibilities to speed up the fitting process, using the implementation of the \textsc{Frankenstein}-code \citep{JenningsEA20} and we avoid fitting the visibilities with large scatter, generally occurring at larger values of k$\lambda$. For BP~Tau, DL~Tau, DR~Tau, Sz~98, and V1094~Sco we limit the visibilities to $\leq$2000 k$\lambda$, whereas we limit the visibilities of CI~Tau and GW~Lup to $\leq$3000 k$\lambda$. Finally, the visibilities of IQ~Tau are limited to $\leq$1500 k$\lambda$. In addition, the visibilities are binned to widths of 1-2$\times$10$^{4}$ $\lambda$; that is, the binned visibilities have widths of 1-2$\times$10$^{4}$ $\lambda$ or 10-20 k$\lambda$. We fit the visibilities using the Markov Chain Monte Carlo (MCMC) implementation of the \textsc{emcee}-package \citep{emcee}. The fitting was performed two-fold, similarly to the approach carried out by \citet{ref:18LoPiHe}: in the first round, we used 2500 iterations to explore the prior space, identifying the best-fit parameters. In the second round, we used 10,000 iterations to explore a confined parameter space around these best-fit parameters. From this second round, we use a final 5000 iterations to identify the median values of the posterior distributions. The lower and upper uncertainties are, respectively, taken to be the 16$^\textnormal{th}$ and 84$^\textnormal{th}$ percentiles. Our fitting results are listed in Table \ref{tab:FitResults} and the fits to the visibilities are shown in Figure \ref{fig:VisProfs}.

\begin{table}[ht!]
    \centering
    \caption{Median values from the fitting posterior distributions and the corresponding uncertainties.}
    \begin{tabular}{c c c c}
        \hline
        Source & a$_i$ & $\sigma_i$ & $\rho_i$ \\
        & [a.u.] & ["] & [k$\lambda]$ \\
        \hline\hline
        BP~Tau & 104$^{+5}_{-10}$ & 0.17$^{+0.04}_{-0.02}$ & - \\
               & 17$^{+39}_{-14}$ & 0.28$^{+0.07}_{-0.04}$ & 251$^{+35}_{-37}$ \\
               & 2$\pm6$ & 0.48$\pm0.07$ & 818$^{+58}_{-66}$ \\
               & 13$^{+7}_{-6}$ & 0.21$^{+0.06}_{-0.05}$ & 904$^{+62}_{-38}$ \\
               & -17$^{+6}_{-7}$ & 0.07$\pm0.03$ & 719$^{+62}_{-67}$ \\
        \hline
        CI~Tau & 186$^{+20}_{-15}$ & 0.46$\pm0.02$ & - \\
               & 135$^{+35}_{-26}$ & 0.43$\pm0.04$ & 161$^{+8}_{-9}$ \\
               & 45$^{+21}_{-14}$ & 0.45$\pm0.06$ & 388$\pm14$ \\
               & 120$^{+24}_{-29}$ & 0.08$^{+0.03}_{-0.01}$ & 516$^{+66}_{-61}$ \\
               & 28$^{+22}_{-16}$ & 0.29$^{+0.07}_{-0.05}$ & 544$^{+42}_{-43}$ \\
               & 111$^{+23}_{-26}$ & 0.17$\pm0.03$ & 1018$^{+26}_{-25}$ \\
               & 63$^{+16}_{-29}$ & 0.16$^{+0.07}_{-0.04}$ & 1401$^{+63}_{-43}$ \\
               & 27$^{+18}_{-15}$ & 0.22$^{+0.05}_{-0.04}$ & 1714$^{+53}_{-55}$ \\
        \hline
        DL~Tau & 251$\pm10$ & 0.39$\pm0.01$ & - \\
               & 208$^{+28}_{-20}$ & 0.33$\pm0.01$ & 215$\pm3$ \\
               & 172$\pm4$ & 0.11$^{+0.02}_{-0.01}$ & 527$^{+11}_{-13}$ \\
               & 48$^{+19}_{-11}$ & 0.39$^{+0.06}_{-0.05}$ & 853$\pm6$ \\
               & 92$\pm9$ & 0.30$\pm0.03$ & 1194$^{+10}_{-12}$ \\
               & 77$^{+7}_{-9}$ & 0.49$^{+0.03}_{-0.02}$ & 1371$\pm4$ \\
               & 99$\pm5$ & 0.22$\pm0.01$ & 1731$^{+7}_{-5}$ \\
        \hline
        DR~Tau & 461$^{+17}_{-21}$ & 0.16$\pm0.01$ & - \\
               & 179$^{+60}_{-65}$ & 0.21$^{+0.02}_{-0.01}$ & 329$^{+21}_{-20}$ \\
               & 210$\pm8$ & 0.09$\pm0.01$ & 807$^{+28}_{-30}$ \\
               & 99$\pm7$ & 0.16$^{+0.01}_{-0.02}$ & 1692$\pm16$ \\
        \hline
        GW~Lup & 100$^{+30}_{-22}$ & 0.31$\pm0.02$ & - \\
               & 106$^{+11}_{-14}$ & 0.18$^{+0.04}_{-0.05}$ & 208$^{+36}_{-56}$ \\
               & 62$^{+8}_{-9}$ & 0.14$^{+0.04}_{-0.03}$ & 688$^{+49}_{-58}$ \\
               & 26$\pm8$ & 0.36$^{+0.06}_{-0.05}$ & 1113$\pm22$ \\
               & 25$\pm6$ & 0.17$\pm0.04$ & 1414$^{+47}_{-49}$ \\
               & 12$\pm5$ & 0.42$^{+0.07}_{-0.06}$ & 1819$^{+31}_{-32}$ \\
        \hline
        IQ~Tau & 69$^{+28}_{-20}$ & 0.36$\pm0.03$ & - \\
               & 64$^{+15}_{-18}$ & 0.26$^{+0.06}_{-0.03}$ & 135$^{+54}_{-31}$ \\
               & 23$\pm4$ & 0.34$\pm0.06$ & 528$^{+17}_{-16}$ \\
               & 11$^{+5}_{-4}$ & 0.33$^{+0.07}_{-0.06}$ & 1013$^{+38}_{-45}$ \\
               & 10$\pm4$ & 0.92$\pm0.07$ & 1200$^{+13}_{-16}$ \\
        \hline
        Sz~98 & 143$^{+7}_{-4}$ & 0.40$\pm0.01$ & - \\
              & 81$^{+9}_{-8}$ & 0.42$^{+0.02}_{-0.03}$ & 216$^{+3}_{-4}$ \\
              & 31$^{+17}_{-7}$ & 0.16$^{+0.04}_{-0.03}$ & 661$^{+30}_{-29}$ \\
              & 16$\pm4$ & 0.55$^{+0.07}_{-0.05}$ & 994$^{+12}_{-13}$ \\
              & 29$^{+4}_{-3}$ & 0.45$^{+0.05}_{-0.04}$ & 1258$^{+9}_{-8}$ \\
              & 104$^{+12}_{-15}$ & 0.04$\pm0.01$ & 1449$^{+65}_{-54}$ \\
        \hline
        V1094~Sco & 131$^{+11}_{-16}$ & 0.66$^{+0.04}_{-0.03}$ & - \\
                  & 166$\pm6$ & 0.48$^{+0.04}_{-0.05}$ & 124$\pm5$ \\
                  & 61$^{+8}_{-7}$ & 0.68$^{+0.04}_{-0.05}$ & 261$\pm4$ \\
                  & 57$^{+6}_{-11}$ & 0.38$\pm0.03$ & 430$\pm5$ \\
                  & 33$\pm4$ & 0.19$\pm0.07$ & 709$^{+31}_{-35}$ \\
                  & 31$\pm8$ & 0.46$\pm0.05$ & 943$^{+9}_{-8}$ \\
                  & 41$^{+4}_{-5}$ & 0.51$^{+0.05}_{-0.06}$ & 1173$\pm7$ \\
                  & 29$\pm4$ & 0.41$\pm0.06$ & 1368$\pm12$ \\
                  & 16$\pm3$ & 0.36$\pm0.06$ & 1619$^{+17}_{-18}$ \\
        \hline
    \end{tabular}
    \label{tab:FitResults}
    
    \vspace{1ex}

     {\raggedright \textbf{Notes.} Gaussian function fit parameters a$_i$, $\sigma_i$, and $\rho_i$, describing the intensity, width, and spatial frequency, respectively (see Eq.~\ref{eq:gauss}). \par}
     
\end{table}

\subsection{Characterising the millimetre continuum emission}
Using the radial brightness distribution from median values of the posterior distributions, we infer the radial locations of substructures. The locations of rings ($r_\textnormal{r}$) and gaps ($r_\textnormal{g}$) are identified through, respectively, the local maxima and minima present in the profiles. Other structures, such as `shoulders' or `plateaus', visible in the radial profiles are visually identified. To characterise the depth and widths of the identified gaps, we follow the approach of \citet{ref:18HuAnDu}. Here, the gap depth is defined as the ratio of the gap intensity ($I_\textnormal{g}$) and that of the neighbouring ring with the smallest flux contrast ($I_\textnormal{r}$), $\Delta I_\textnormal{g}=I_\textnormal{g}/I_\textnormal{r}$. Given the ratio, the gaps with the smallest value reported for the depth are the deepest ones. The width of the gap is determined as the radial width between the inner and outer edges of the gaps ($r_\textnormal{g,i}$ and $r_\textnormal{g,o}$), which are the radial locations where the intensity equals $I_\textnormal{mean}=0.5\left(I_\textnormal{g}+I_\textnormal{r}\right)$. Therefore, the width is given by $\Delta r_\textnormal{g}=r_\textnormal{g,o}-r_\textnormal{g,i}$.

We also determine the continuum disc radius ($R_\textnormal{disc}$) using a curve-of-growth method (see also \citealt{ref:16AnWiMa} and/or \citealt{ref:22StHoDi} for more information) for radii containing 68\%, 90\%, and 95\% of the flux density. The values for the disc radii are listed in Table \ref{tab:DiskResults}. In short, we use aperture photometry to investigate at what radius the flux density within the apertures flattens off. We increase the apertures by the pixel size and limit our range to where we see no significant substructures in the ALMA radial profiles, as we are only using high-resolution observations, and therefore we may lack the resolving power for weak outer disc features. \\ 

\begin{figure*}[ht!]
    \centering
    \includegraphics[width=\textwidth]{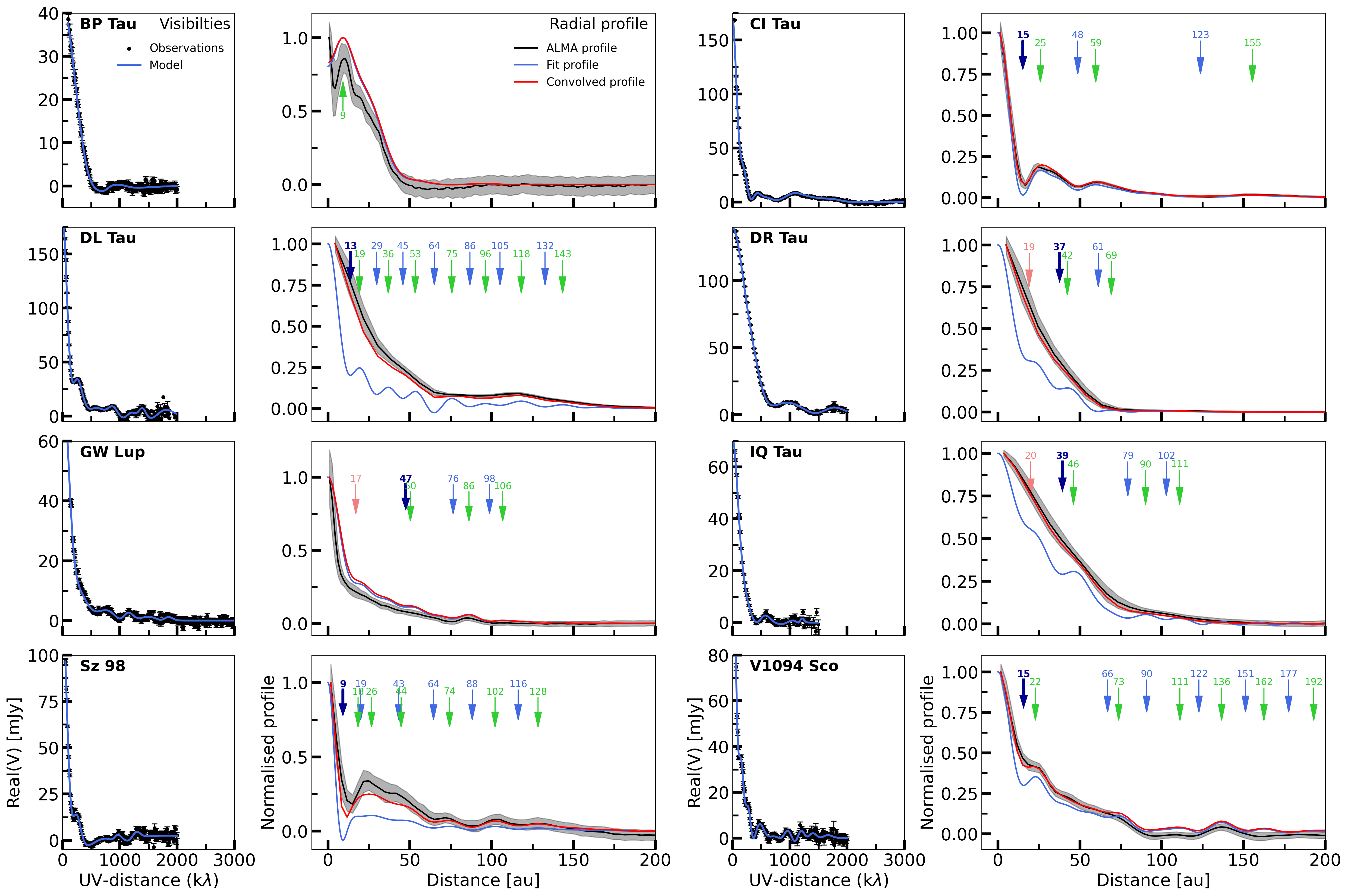} 
    \caption{Fit to the visibilities (blue line in the left panels for each source) and the various radial profiles (right panels): one from the ALMA image (black line), the model radial profile (blue line), and the radial profile from the model convolved with a Gaussian kernel tailored after the resolving beam (red line). The shaded grey area indicates the standard deviation in each radial bin of the ALMA profile. The gaps and rings (or local minima and maxima) and plateaus/shoulders are indicated by, respectively, the blue, green, and red arrows, where the values indicate their integer radial positions. The location of the innermost gap is emphasised by the dark blue arrows.}
    \label{fig:VisProfs}
\end{figure*}

The model radial profiles, together with the identified gaps and rings, are shown in Figure \ref{fig:VisProfs}. Here, we also show the ALMA continuum image and the model radial profile convolved with a Gaussian kernel tailored after the ALMA resolving beam. The convolved radial profiles clearly show that the beam is responsible for smoothing out weak structures in the ALMA image and, subsequently, that these features cannot be distinguished from the images. All the retrieved locations of the gaps and the rings, and the gap widths and depths are listed in Table \ref{tab:DiskResults}. Additionally, in Figure \ref{fig:Images} we display a comparison between the ALMA continuum image, the model image, the model image convolved with a Gaussian kernel, and the residuals. The residuals are obtained by subtracting the convolved model image off the ALMA image. Similarly to what is seen for the radial profiles, the convolution with the Gaussian kernel smooths out the various weak substructures visible in the model and therefore become unidentifiable in the images.

Besides the various gaps and rings, we also visually identified a few plateaus, which may hint at additional substructures. These plateaus are indicated by the red arrows in Figure \ref{fig:VisProfs}. To summarise, we identify three plateaus: in DR~Tau at $\sim$19 au, in GW~Lup at $\sim$17 au, and in IQ~Tau at $\sim$20 au.

Overall we find that the convolved images represent the ALMA images quite well, but there are some residuals that require attention. The residuals of BP~Tau are mostly negative, indicating larger flux in the model image following the identified cavity in the models. The central, unresolved spot in the ALMA image may be attributed to free-free emission or emission from the host star itself (see e.g. \citealt{ref:24RoMeMa}). For CI~Tau, the residuals show that clear asymmetric structures are found in the ALMA image. These residuals show that the outer edges of the gap are not perfectly oval shaped as it is in the model. DL~Tau, DR~Tau, and IQ~Tau all three show clear residuals in the central positions. These asymmetries likely arise from sub-pixel offsets between the peaks of the ALMA images and the model images. We note that the observations for these sources have the lowest spatial resolution ($\sim$0.11"), which provides a potential explanation why these residuals are the largest. Higher resolution observations will likely provide better fits and cleaner residuals. The residuals of GW~Lup are dominated by the model overfitting the width of the inner region, as can also be seen from the radial profiles (see Figure \ref{fig:VisProfs}). Additionally, there may be a sub-pixel offset for the location of the maximum flux in the ALMA image, which causes the asymmetric feature to appear in the residuals. For both Sz~98 and V1094~Sco, we see that the residuals are rather patchy, which may be caused by the use of only the highest resolution observations available. Combining these observations with those of lower resolution, which are better at resolving the largest scales, may improve the residuals.

\begin{table*}[ht!]
    \caption{Identified substructures and dust continuum sizes for the sources.}
    \centering
    \begin{tabular}{c c c c c c c c}
        \hline
        Source & $r_\textnormal{g}$ & $r_\textnormal{r}$ & $\Delta r_\textnormal{g}$ & $\Delta I_\textnormal{g}$ & \multicolumn{3}{c}{$R_\textnormal{disc}$ (Model/ALMA)$^{(a)}$} \\
               & [au] & [au] & [au] & & 68\% [au] & 90\% [au] & 95\% [au] \\
        \hline\hline
        BP~Tau & - & 9 & - & - & 30.0/29.0 & 40.6/38.1 & 46.7/44.7 \\
        \hline
        CI~Tau & \textbf{15.2} & 25.9 & 7.8 & 0.10 & 105.3/114.2 & 172.5/180.3 & 189.3/199.4 \\
               & 48.7 & 59.7 & 9.4 & 0.59 & & & \\
               & 123.7 & 155.5 & 27.2 & 0.59 & & & \\
        \hline
        DL~Tau & \textbf{13.8} & 19.2 & 4.5 & 0.81 & 112.0/115.2 & 147.2/156.8 & 163.2/176.0 \\
               & 29.9 & 37.0 & 5.9 & 0.65 & & & \\
               & 45.9 & 53.3 & 7.0 & 0.59 & & & \\
               & 65.0 & 75.9 & 9.6 & -0.44 & & & \\
               & 86.7 & 96.2 & 7.7 & 0.38 & & & \\
               & 105.0 & 117.9 & 8.2 & 0.66 & & & \\
               & 132.6 & 143.2 & 9.3 & 0.46 & & & \\ 
        \hline
        DR~Tau & \textbf{37.7} & 42.3 & 3.9 & 0.96 & 39.0/42.9 & 54.6/66.3 & 89.7/97.5  \\
               & 61.2 & 69.2 & 6.0 & 0.33 & & & \\
        \hline
        GW~Lup & \textbf{47.5} & 50.4 & 2.4 & 0.99 & 67.4/65.1 & 102.3/110.8 & 115.5/134.1 \\
               & 76.5 & 86.1 & 11.0 & 0.56 & & & \\
               & 98.7 & 106.8 & 6.5 & 0.54 & & & \\
        \hline
        IQ~Tau & \textbf{39.3} & 46.1 & 5.7 & 0.95 & 58.0/68.6 & 103.0/118.8 & 118.8/145.2 \\
               & 79.3 & 90.2 & 10.2 & 0.62 & & & \\
               & 102.8 & 111.0 & 7.5 & 0.73 & & & \\
        \hline
        Sz~98 & \textbf{8.6} & 16.4 & 4.6 & -0.72 & 109.8/107.3 & 147.3/146.0 & 159.7/162.2 \\
              & 19.3 & 27.3 & 3.1 & 0.97 & & & \\
              & 42.8 & 45.3 & 2.1 & 1.00 & & & \\
              & 64.3 & 74.3 & 8.9 & 0.63 & & & \\
              & 88.1 & 102.0 & 13.3 & 0.21 & & & \\
              & 116.0 & 128.0 & 10.5 & 0.61 & & & \\
        \hline
        V1094~Sco & \textbf{15.6} & 22.8 & 5.8 & 0.85 & 162.4/131.4 & 239.3/238.1 & 255.4/256.7 \\
                  & 67.0 & 73.7 & 6.4 & 0.92 & & & \\
                  & 90.8 & 111.2 & 14.3 & 0.40 & & & \\
                  & 122.8 & 136.7 & 9.8 & 0.22 & & & \\
                  & 151.2 & 162.5 & 9.5 & 0.37 & & & \\
                  & 177.6 & 193.0 & 12.0 & 0.37 & & & \\
                  & 200.2 & 217.2 & 6.8 & 0.95 & & & \\
        \hline
    \end{tabular}
    \label{tab:DiskResults}
    
    \vspace{1ex}

     {\raggedright \textbf{Notes.} $^{(a)}$We list both the values for the outer dust disc radii retrieved from the model image and from the ALMA image. The value before the slash is obtained from the model images, whereas the value after the slash is obtained from the ALMA image. $r_\text{g}$, $r_\text{r}$, $\Delta r_\text{g}$, and $\Delta I_\text{g}$ are the gap location, ring location, gap width and depth, respectively. The values used as $R_\text{gap}$ are highlighted in boldface. \par}

\end{table*}

\begin{figure*}[ht!]
    \centering
    \includegraphics[width=\textwidth]{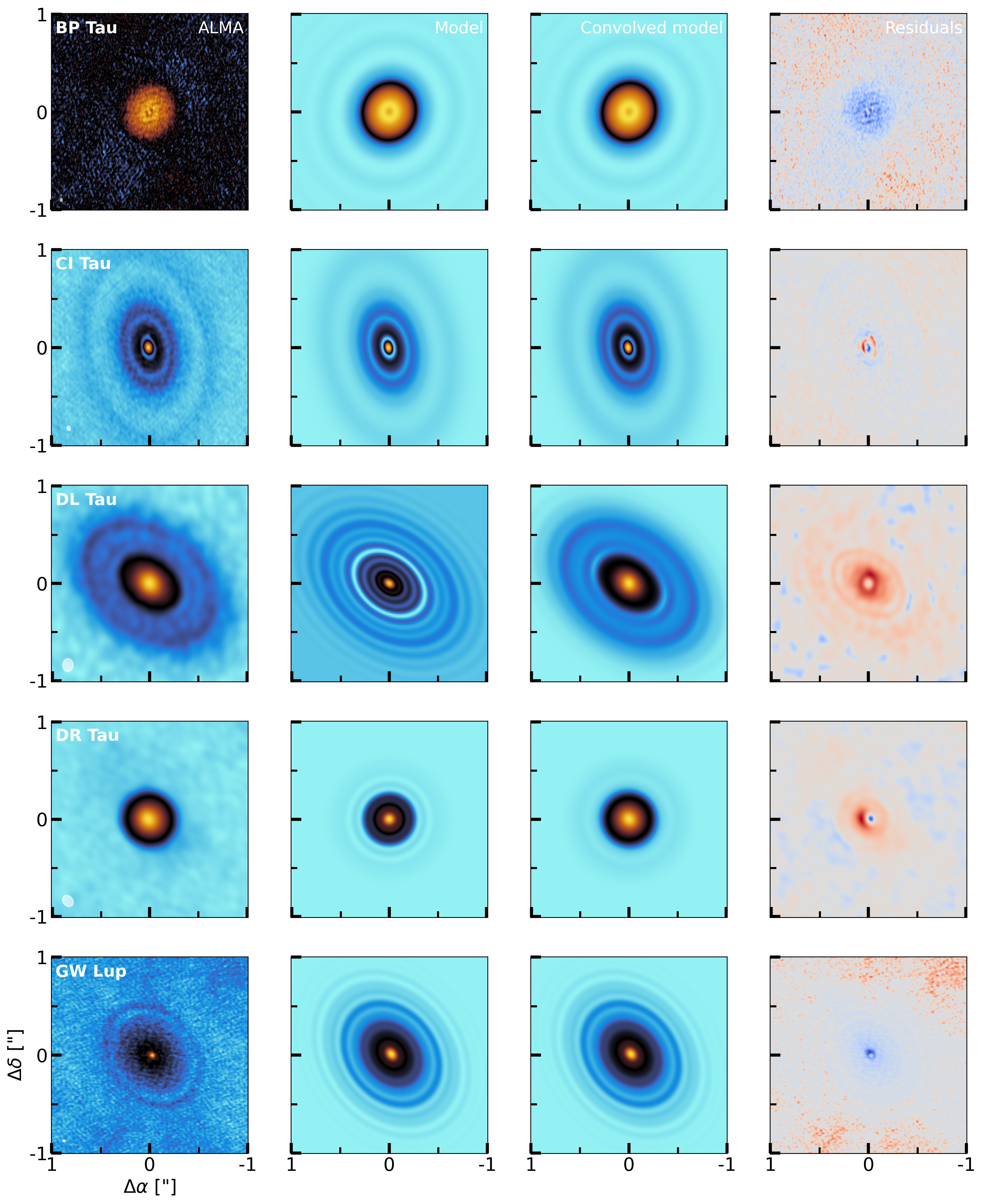}
    \caption{Comparison between the ALMA image (first column), the model image (second column), the convolved model (third column), and the residuals (final column). All images, except the residuals, are shown with an arcsinh-stretch colourmap. The residuals are taken by subtracting the convolved model image off the ALMA image. Here, red residuals indicate more flux in the ALMA image, whereas blue residuals indicate more flux in the convolved model image.}
    \label{fig:Images}
\end{figure*}
\begin{figure*}[ht!]
    \centering
    \ContinuedFloat
    \includegraphics[width=\textwidth]{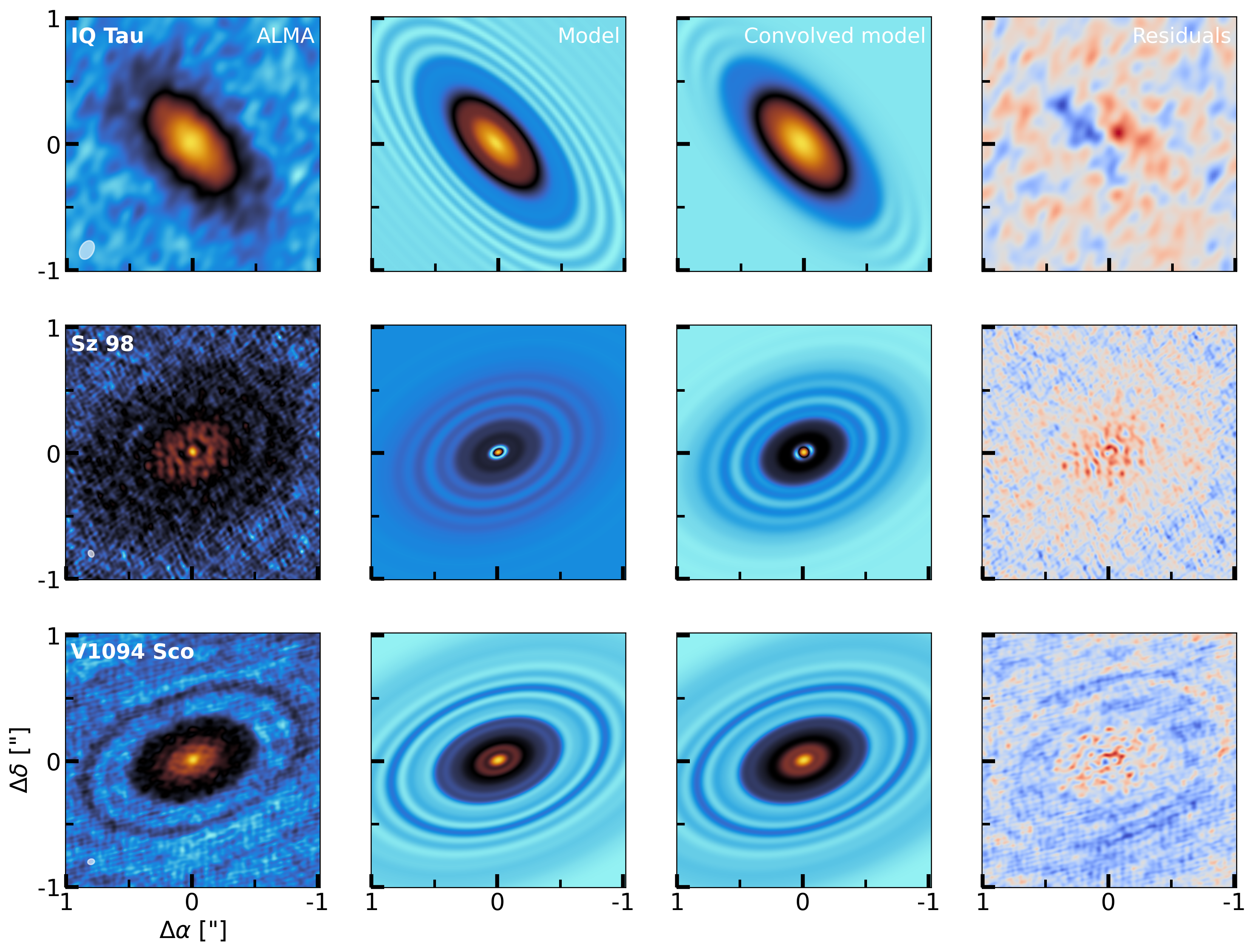}
    \caption{Continuation of Fig.~\ref{fig:Images}.}
\end{figure*}

\subsection{Comparison with previous works}
Almost all of the discs, except Sz~98, have been modelled before, we compare the inferred substructures:

\textit{BP Tau}: BP~Tau has previously been fitted by \citet{ref:23ZhKaLo}, but no fits have been carried out for the high-resolution observations used in this work. We note that the models of \citet{ref:23ZhKaLo} already point towards a small cavity in this disc, which now seems confirmed by our results. Additional analysis of this source will be presented in a future publication of the PI's team.

\textit{CI Tau}: The high-resolution CI~Tau observations used in this work have previously been modelled by \citet{ref:18ClTaJu} and \citet{ref:21RoIlFa}. They find radial locations for the gaps of $\sim$12, $\sim$45, and $\sim$114 au, and radii of $\sim$23, $\sim$54, and $\sim$114 au for the rings. Generally, these values are close to our values, except for the outer most ring, for which we find a larger value of $\sim$155 au.

\textit{DL Tau}: DL~Tau has been modelled by \citet{ref:23ZhKaLo} using the same technique, but also has been given an extensive treatment by \citet{ref:22JeTaCl}. Both works yield similar radial brightness profiles. In addition, \citet{ref:23ZhKaLo} lists gap locations of $\sim$15, $\sim$31, $\sim$47, $\sim$66, $\sim$95, and $\sim$129 au, which all fall within 10 au of our listed values. The fit by \citet{ref:23ZhKaLo} suggests deeper gaps. As is seen in Figure \ref{fig:VisProfs} our convolved profile does not fully match the ALMA profile, suggesting that our modelled rings are not as strong as those of \citet{ref:23ZhKaLo}, yielding smaller gap depths.

\textit{DR Tau}: Similar to DL~Tau, DR~Tau has been modelled by \citet{ref:22JeTaCl} and \citet{ref:23ZhKaLo}. Again, we find that the respective brightness profiles are rather similar, except we note that we find an additional potential substructure gap-ring combination at, respectively, 61.2 and 69.2 au. The other gaps are found at similar locations, although we find that the first gap identified by \citet{ref:23ZhKaLo} ($\sim$18 au) appears as a plateau in our profile.

\textit{GW Lup}: GW~Lup was modelled by \citet{ref:22JeBoTa}, which finds a similar looking brightness profile, with a plateau in the inner regions ($<$0.1"). Overall, we find a few more potential substructures apart from the well-known ring at 86 au. As these substructures are not identified by \citet{ref:22JeBoTa}, we propose to treat these potential substructures with caution.

\textit{IQ Tau}: IQ Tau has also been modelled by \citet{ref:23ZhKaLo}. We find the same location for the gap-ring combination around $\sim$40 au. However, the outer two rings identified in our work are not found in their work. Therefore, we propose to treat these two potential substructures with caution. Higher spatial resolution observations may proof if these substructures do indeed exist or not.

\textit{V1094 Sco}: These higher resolution observations of V1094~Sco have not been modelled before, but \citet{ref:18TeDiAn} modelled lower resolution ($\sim$0.18") observations. They only provided locations for gaps located at 100 and 170 au. Our fit to the higher resolution observations thus provide hints for potentially more substructures, with the most notable gap at $\sim$16 au.

\end{appendix}




\end{document}